\documentclass[a4paper,11pt]{article}
\pdfoutput=1 % if your are submitting a pdflatex (i.e. if you have
\RequirePackage{snapshot}  % required by bundledoc

\usepackage{jheppub} % for details on the use of the package, please
\usepackage{bm,amsmath,amssymb,slashed,graphicx,%
            enumerate,alltt,xspace,multirow,xcolor,mathrsfs}
\usepackage{fancyvrb}
\usepackage{booktabs}
\usepackage{graphicx}
\usepackage{subcaption}
\usepackage[utf8]{inputenc}
\usepackage[export]{adjustbox}
\usepackage{textcomp}

\makeatletter
\g@addto@macro\bfseries{\boldmath}
\makeatother

\definecolor{labelkey}{rgb}{0,0.5,0.0}

\usepackage{listings}
\definecolor{darkgreen}{rgb}{0,0.7,0}
\definecolor{ddarkgreen}{rgb}{0,0.5,0}
\definecolor{grey}{rgb}{0.5,0.5,0.5}
\definecolor{orange}{rgb}{1.0,0.4,0.4}
\definecolor{cyan}{rgb}{0.0,1.0,1.0}
\definecolor{magenta}{rgb}{1.0,0.0,1.0}
\title{Quarks and gluons in the Lund plane}

\author[a]{Frederic Dreyer,}%
\author[b]{Gregory Soyez,}%
\author[c]{Adam Takacs}%

\emailAdd{frederic.dreyer@physics.ox.ac.uk}
\emailAdd{gregory.soyez@ipht.fr}
\emailAdd{adam.takacs@uib.no}

\affiliation[a]{Rudolf Peierls Centre for Theoretical Physics, Parks Road, Oxford OX1 3PU, UK}
\affiliation[b]{IPhT, Universit\'{e} Paris-Saclay, CNRS UMR 3681, CEA Saclay, F-91191 Gif-sur-Yvette, France}
\affiliation[c]{Department of Physics and Technology, University of Bergen, 5007 Bergen, Norway}

\date{Received: date / Accepted: \today}
\abstract{
  Discriminating quark and gluon jets is a long-standing topic in
  collider phenomenology.
  In this paper, we address this question using the Lund jet plane
  substructure technique introduced in recent years.
  We present two complementary approaches: one where the quark/gluon
  likelihood ratio is computed analytically, to single-logarithmic
  accuracy, in perturbative QCD, and one where the Lund declusterings
  are used to train a neural network.
  For both approaches, we either consider only the primary
  Lund plane or the full clustering tree. 
  The analytic and machine-learning discriminants are shown to be
  equivalent on a toy event sample resumming exactly leading collinear
  single logarithms, where the analytic calculation corresponds to the
  exact likelihood ratio.
  On a full Monte Carlo event sample, both approaches show a good
  discriminating power, with the machine-learning models usually being
  superior.
  We carry on a study in the asymptotic limit of large logarithm,
  allowing us to gain confidence that this superior performance comes
  from effects that are subleading in our analytic approach.
  We then compare our approach to other quark-gluon discriminants in
  the literature.
  Finally, we study the resilience of our quark-gluon discriminants
  against the details of the event sample and observe that the
  analytic and machine-learning approaches show similar behaviour.
}

\begin{document}

\maketitle

%\tableofcontents 

%======================================================================
\section{Introduction}\label{sec:intro}

Quarks and gluons, the constituents of the proton, are fundamental
entities of essential relevance for physics at the Large Hadron
Collider (LHC) at CERN.
While these particles are ubiquitous at hadron colliders, they are
never observed directly, but rather fragment and hadronise immediately
into collimated sprays of colourless hadrons.
These decay products are referred to as jets and are generally
defined through the application of a sequential recombination
algorithm (see e.g.~\cite{Salam:2009jx,Marzani:2019hun}).

As most of the jets observed at collider experiments arise from the
fragmentation of a light parton, a detailed understanding of their
properties is crucial for experimental analyses.
In this context, many experimental studies make use of tools that
reliably identify the flavour of jets, e.g.\ to enhance signals
from new physics (decaying predominantly to quarks) from QCD
backgrounds (producing predominantly gluon jets).
Since quark and gluon branch into one another, it is highly
non-trivial to even define what is meant by a ``quark jet'' or a
``gluon jet'' (see, for example, the discussion in
Ref.~\cite{Gras:2017jty}, as well as
Refs.~\cite{Komiske:2018vkc,Larkoski:2019nwj}).
As a direct consequence, it is delicate to introduce a
properly-defined flavoured jet algorithm~\cite{Banfi:2006hf}.

Over the past decade, jet substructure, the study of the internal
dynamics of jets, has proven a useful approach to study the decay of
heavy particles at and above the electroweak scale, providing a
promising avenue to search for signs of new physics beyond the
Standard Model (see
\cite{Larkoski:2017jix,Asquith:2018igt,Marzani:2019hun} for recent
reviews).
While jet substructure has applications in many directions
including for example precision measurements in QCD and the study of
the quark-gluon plasma produced in heavy-ion collisions, recent
years have seen an increasing interest in leveraging progress in deep
learning to a range of jet tagging
problems~\cite{jet_images2,jets_w,deep_top1,jets_comparison,particlenet,jedinet,DeepJet,Du:2021pqa,Du:2020pmp,Apolinario:2021olp}.

Several jet substructure techniques have been introduced to address
the question of quark/gluon discrimination. This includes jet-shape
based observables like jet
angularities~\cite{Berger:2003iw,Almeida:2008yp}, energy-energy
correlation functions~\cite{Larkoski:2013eya} or the jet
charge~\cite{Field:1977fa,Krohn:2012fg,Kang:2021ryr}, counting
observables like the charged track multiplicity or the Iterative Soft
Drop multiplicity~\cite{Frye:2017yrw}, as well as a series of recent
deep-learning-based approaches using a range of network architectures
and
inputs~\cite{Louppe:2017ipp,Komiske:2018cqr,Qu:2019gqs,Lee:2019ssx,Dreyer:2020brq}.
Other techniques, such as jet
topics~\cite{Metodiev:2018ftz,Komiske:2018vkc,Brewer:2020och}, are
based on a statistical ensemble of events and are directly meant to
obtain separate distributions for quarks and gluons. These are not
discussed here as we instead target quark/gluon discriminants working
on individual jets.

Recently, the Lund Jet Plane has been introduced~\cite{Dreyer:2018nbf}
as a powerful technique to tackle a wide range of jet substructure
applications. For example, the primary Lund plane density has been
measured by the ATLAS~\cite{Aad:2020zcn} and
ALICE~\cite{ALICE-PUBLIC-2021-002} collaborations highlighting, for
example, differences between general-purpose Monte-Carlo event
generators.
This Lund plane density is amenable to precision calculations in
perturbative QCD~\cite{Lifson:2020gua}, showing an agreement with the
ATLAS measurement.
Finally, Lund-plane variables can be used as inputs to
machine-learning tagger~\cite{Dreyer:2018nbf,Dreyer:2020brq}.

In this paper, we will use the Lund plane approach to study
quark/gluon discrimination.
We will do this using both an analytic approach and machine-learning
tagging methods.
In both cases, we will build two taggers: one based on information
from primary Lund declusterings only, and a second based on the full
Lund declustering tree.
Our analytic approach is based on a resummed calculation of the
likelihood ratio at the single-logarithmic accuracy, i.e.\ matching
the logarithmic accuracy obtained in Ref.~\cite{Lifson:2020gua} for
the primary Lund plane density.
We note that likelihood ratios have already been relied upon in the
context of boosted-jet discrimination, for example, shower
deconstruction~\cite{Soper:2011cr,Soper:2012pb,FerreiradeLima:2016gcz}.
Our machine-learning taggers follow the guidelines from
Refs.~\cite{Dreyer:2018nbf,Dreyer:2020brq}.
One of the main novelties of this work is that we will aim to gain a
first-principles understanding of the behaviour of the neural network
by comparing it with our analytic discriminants in specific limits
where the analytic approach is known to be optimal.
This provides cross-validation of both approaches, shedding light on
the importance of subleading effects in the analytic tagger and
providing information on the convergence of deep-learning methods.

The paper is organised as follows. 
In section~\ref{sec:earlier} we describe the Lund plane which will
serve as framework for this study.
We describe the analytic strongly angular-ordered Lund-plane discriminant in
section~\ref{sec:discrim-analytic}, both for the primary Lund plane
and for the full clustering tree.
We discuss the inclusion of clustering logarithms in
section~\ref{sec:clustering-logs}.
The baseline machine learning models used in our comparisons are
described in section~\ref{sec:ml-approaches}, and we provide a
validation of the analytic discriminants against these models using a
toy shower in section~\ref{sec:mc-collinear}.
Finally, we perform an in-depth comparison of the performance and
resilience of a wide range of methods on full Monte Carlo simulations
in section~\ref{sec:mc-full}, showing that our approaches are either
better or on par with state-of-the-art methods both in terms of
discriminating power and in terms of resilience.

The code implementing our analytic quark-gluon discriminant based on
Lund declusterings is available at
\href{https://gitlab.com/gsoyez/lund-quark-gluon}{https://gitlab.com/gsoyez/lund-quark-gluon}.

%======================================================================
\section{Lund plane(s) and baseline discriminants}\label{sec:earlier}

In order to fix once and for all the notations to be used throughout
this paper, we briefly remind the reader of how the primary Lund plane
declusterings are constructed~\cite{Dreyer:2018nbf}. We also introduce
a generalisation beyond the primary plane that instead keeps the full
{\em declustering tree} that we exploit later in this paper.

\paragraph{Primary Lund declusterings.}
For a given jet, we first recluster its constituents with the
Cambridge/Aachen
algorithm~\cite{Dokshitzer:1997in,Wobisch:1998wt}.\footnote{The reason
  why we use the Cambridge/Aachen algorithm instead of other
  algorithms of the generalised-$k_t$ family~\cite{Cacciari:2008gp}
  algorithm, is discussed in section 2.4 of
  Ref.~\cite{Dreyer:2018nbf}.} We then build the list of primary
declusterings as follows:
\begin{enumerate}
\item start with $j$ being the full reclustered jet;
\item undo the last step of the clustering, $j\rightarrow j_1+j_2$,
  giving two subjets $j_1$ and $j_2$. We assume without loss of
  generality that $j_1$ is the ``harder branch'' i.e.\ that $p_{t1}>
  p_{t2}$.
\item We define the set of coordinates $\mathcal{T}=\{\Delta, k_t,
  z,\psi,\dots\}$ for this branching:
  \begin{align}
    \Delta & = \sqrt{(y_1-y_2)^2 + (\phi_1-\phi_2)^2},&
       k_t & = p_{t2}\Delta,  \nonumber\\
       z   & = \frac{p_{t2}}{p_{t1}+p_{t2}},&
      \psi &= \tan^{-1}\frac{y_2-y_1}{\phi_2-\phi_1}.
     \label{eq:lund-coordinates}
  \end{align}
\item Iterate by going back to step~2 with $j\leftarrow
  j_1$.
\end{enumerate}
This produces a tuple, ordered from the first declustering to the last,
\begin{equation}\label{eq:lund-primary-tuple}
  \mathcal{L}_\text{primary} = \left[\mathcal{T}_1,\dots,\mathcal{T}_i,\dots,\mathcal{T}_n \right]
\end{equation}
that we refer to as the primary Lund declusterings associated with the
jet $j$.

The averaged primary Lund plane density is then simply defined as the
average number of declusterings for a given $\ln\Delta$ and $\ln k_t$:
\begin{equation}\label{eq:average-lund-density}
  \rho(\Delta,k_t) =
  \frac{1}{N_\text{jets}}\frac{d^2N}{d\ln\Delta\,d\ln k_t}.
\end{equation}
This quantity has been measured by the ATLAS collaboration in
Ref.~\cite{Aad:2020zcn} (see also Ref.~\cite{Ehlers:2021njb} for a
preliminary ALICE measurement) and studied analytically in
Ref.~\cite{Lifson:2020gua}.

\paragraph{Lund declustering tree.}
Instead of focusing only on following the hardest branch through the
declustering, one can retain the full Cambridge/Aachen tree structure
yielding an associated tree of Lund variables:
\begin{equation}\label{eq:lund-tree-raw}
  \mathcal{L}_\text{tree}(j)
  = \left[\left(
      \mathcal{T}(j),
      \mathcal{L}_\text{tree}(j_1),
      \mathcal{L}_\text{tree}(j_2)
    \right) \right]
\end{equation}
where the tree, $\mathcal{L}_\text{tree}(j)$, associated with a jet
$j$ has a set $\mathcal{T}(j)$ of Lund coordinates associated with the
branching $j\to j_1+j_2$, with $p_{t1}>p_{t2}$, as well as sub-trees
$\mathcal{L}_\text{tree}(j_1)$ and $\mathcal{L}_\text{tree}(j_2)$
associated with $j_1$ and $j_2$ respectively.
Note that this structure can be flattened into a tuple
\begin{equation}\label{eq:lund-tree-flat}
  \mathcal{L}_\text{tree}(j)
  = \left[
    \left(\mathcal{T}_1,i_{\text{hard},1}, i_{\text{soft},1}\right),
    \dots,
    \left(\mathcal{T}_i,i_{\text{hard},i}, i_{\text{soft},i}\right),
    \dots,
    \left(\mathcal{T}_n,i_{\text{hard},n}, i_{\text{soft},n}\right)
  \right],
\end{equation}
where $i_{\text{hard},i}$ (resp.\ $i_{\text{soft},i}$) indicate the
index in the tuple for the next branching along the hard (resp.\ soft)
branch, or 0 in the lack thereof.

\paragraph{Iterated Soft Drop multiplicity.}
The iterated Soft Drop multiplicity~\cite{Frye:2017yrw} can be
straightforwardly defined from the tuple of primary Lund declusterings
as the the number of declusterings satisfying a given Soft
Drop~\cite{Larkoski:2014wba,Dasgupta:2013ihk} condition
$z_i>z_\text{cut}(\Delta_i/R)^\beta$ with $R$ the jet radius and
$z_\text{cut}$ and $\beta$ the Soft Drop parameters, with $\beta<0$.
A standard choice is to take $\beta=-1$ so as to effectively use Soft
Drop to impose a cut on $z\Delta$ which is similar to a $k_t$ scale.
In this paper, we define the iterated Soft Drop multiplicity as
the number of primary declusterings above a fixed $k_t$ cut.
This is motivated by the fact that a dimensionful $k_t$ cut is more
adequate than a cut on $z\Delta$ to separate between a perturbative
region ($k_t>k_{t,\text{cut}}$) and a non-perturbative region
($k_t<k_{t,\text{cut}}$).\footnote{The fundamental physics motivation
  behind this choice is that, in a resummed calculation in
  perturbative QCD, the scale entering the strong coupling is
  typically the $k_t$ of the emission.}
The use of a $k_t$ cut will also be used with all the other methods
introduced in this paper, allowing for a direct comparison of their
performance.

Analytically, one can show that the Iterated Soft Drop multiplicity is
the optimal quark-gluon discriminant in the double-logarithmic
approximation (see below for a proof).

\paragraph{Baseline discriminant: the average Lund-plane density.}\label{sec:average-primary-density}
The baseline approach we will consider throughout this paper is the
one that was introduced in Ref.~\cite{Dreyer:2018nbf}.
We first compute the average primary Lund plane densities
$\rho_{q,g}(\Delta,k_t)$ separately for the quarks and gluons samples,
respectively. For a given jet with Lund declusterings
$\{(\Delta_i,k_{t,i})\}$, we then define a likelihood ratio
\begin{equation}\label{eq:likelihood-density}
  \mathbb{L}_\text{density} = \prod_i \frac{\rho_g(\Delta_i,k_{t,i})}{\rho_q(\Delta_i,k_{t,i})} \,.
\end{equation}
In practice, the average densities $\rho_{q,g}(\Delta,k_t)$ are
computed in bins of $\ln\Delta$ and $\ln k_t$.

This approach has already shown to give good results when applied to
discriminating boosted $W$ bosons from QCD jets. The application to
quark-gluon discrimination considered in this paper is actually
simpler as it does not involve a separate isolation and treatment of a
hard two-prong decay as was the case in~\cite{Dreyer:2018nbf} for
boosted $W$ tagging.

Eq.~(\ref{eq:likelihood-density}) would be an optimal discriminant if
the Lund plane declusterings were independent. We know that this is
not the case in practice due to effects such as the energy lost by the
leading parton, flavour changes and clustering effects, as discussed
in~\cite{Lifson:2020gua}.
However, we still expect this approach to yield a better performance
than the Iterated Soft Drop multiplicity as it captures some aspects
of the quark and gluon radiation patterns beyond the soft-collinear
approximation.
Throughout this paper, we will consider quark-gluon tagging
using the average Lund plane density as a reference approach from
which we want to build more performant discriminants.

%======================================================================
\section{Analytic discriminants in the Lund-plane with strong angular ordering}\label{sec:discrim-analytic}

\subsection{Generic considerations}

In this section, we introduce a series of quark-gluon discriminants
based on a first-principles treatment of the Lund plane declusterings
in perturbative QCD.
The performance of these new tools will be assessed later in
sections~\ref{sec:mc-collinear} and~\ref{sec:mc-full}.

The core idea is to explicitly compute the likelihood ratio
\begin{equation}\label{eq:likelihood-ratio}
  \mathbb{L} = \frac{p(g|\mathcal{L})}{p(q|\mathcal{L})}\, ,
\end{equation}
for a set of Lund declusterings $\mathcal{L}$ --- either primary-only
or including the full tree --- where $p(q,g|\mathcal{L})$ denotes the
probability for the jet to come from either a quark (or
anti-quark)\footnote{Throughout this paper, we do not distinguish
  between quarks and anti-quarks.} or a gluon for the given set of
declusterings.
Note that for a fixed composition of initial quarks and gluons, we can
alternatively compute
\begin{equation}\label{eq:likelihood-ratio-base}
  \mathbb{L} = \frac{p_g(\mathcal{L})}{p_q(\mathcal{L})}\, ,
\end{equation}
where $p_{q,g}(\mathcal{L})$ is the probability to observe the given
set of Lund declusterings assuming the jet is either a quark or a
gluon.
For this procedure to be infrared-and-collinear safe, we only consider
emissions above a given (relative) transverse momentum cut, i.e.\
require $k_t\ge k_{t,\text{cut}}$.

In the (double-logarithmic) soft-collinear limit, emissions are
independent and the single-emission probability for quarks and gluons
only differ by the overall colour factor ($C_F$ for a quark, $C_A$ for
a gluon). For $n$ primary emissions $\mathcal{T}_i$, one therefore has
\begin{equation}\label{eq:likelihood-ll}
  \mathbb{L}_{LL} = \prod_i
  \frac{p_g(\mathcal{T}_i)}{p_q(\mathcal{T}_i)} = \left(\frac{C_A}{C_F}\right)^n.
\end{equation}
This shows that the likelihood ratio is only a (monotonic) function of
$n$ and hence that the iterated Soft Drop multiplicity is the optimal
discriminant at leading (double) logarithmic accuracy. 
In this limit, additional, non-primary, declusterings in the full
Lund tree all come with a factor $C_A$ and therefore do not
contribute to Eq.~\eqref{eq:likelihood-ll}.

In what follows, we want to extend this result to single-logarithmic
accuracy, as what was done in~\cite{Lifson:2020gua} for the average
primary Lund plane density.
For this, several single-logarithmic effects have to be taken into
account: (i) corrections to the running of the strong coupling, (ii)
collinear effects stemming either from splittings where the flavour of
the leading parton changes or from finite $z$ splittings, (iii)
clustering effects where the exact Cambridge/Aachen clustering has to
be taken into account for multiple soft emissions at commensurate
angles, and (iv) effects of multiple soft emissions at large angles.

In the (soft-collinear) limit where emissions are independent, running
coupling effects --- (i) in the above list --- do not change the
double-logarithmic result in Eq.~(\ref{eq:likelihood-ll}) as both the
quark and gluon probabilities are multiplied by $\alpha_s$ taken at
the same scale, namely the (relative) $k_t$ of the
emission. Running-coupling corrections will nevertheless appear
together with the other single-logarithmic contributions and this is
discussed below.

Next, the effect of soft-wide-angle radiation --- (iv) in the above
list --- depends on the details of the hard process that underlines
the samples of quark and gluon jets. These contributions would, for
example, be different in ``quark jets'' in $qg\to Zq$ and $qq\to qq$
events (see e.g. the discussion in Ref.~\cite{Gras:2017jty}). In this
paper, we will focus on universal aspects in the collinear limit and
therefore neglect these contributions which scale like the square of
the jet radius. It would nevertheless be interesting, in a follow-up
study, to investigate if the analytic techniques developed in this
paper could be used to assess the process-dependence of quark-gluon
tagging, potentially in combination with the concept of jet
topics~\cite{Metodiev:2018ftz}.

Clustering effects are delicate to handle in an analytic calculation
as even in the large-$N_c$ limit they, in principle, require the full
matrix angular dependence for an arbitrary number of emissions
strongly-ordered in energy. Since we can expect that collinear
effects, and flavour-changing contributions in particular, are
numerically dominant in the context of quark-gluon discrimination, we
will as a first step neglect clustering effects.
In other words, in this section we work in the regime where emissions are
strongly ordered in angle and derive a quark-gluon discriminant either
using only primary emissions, section~\ref{sec:collinear}, or using
the full clustering tree, section~\ref{sec:tree}.

We come back to the question of clustering logarithms in
section~\ref{sec:clustering-logs}.
We will see explicitly in our Monte Carlo simulations in
section~\ref{sec:mc-full} that clustering effects have a smaller
numerical impact on quark-gluon discrimination that the collinear
enhancements.

%----------------------------------------------------------------------
\subsection{Optimal discriminant for primary Lund declusterings}\label{sec:collinear}

We start by considering only the primary Lund plane declusterings
$\{(\Delta_i,k_{t,i})\}$ with $k_t\ge k_{t,\text{cut}}$. For these, we
want to compute the likelihood ratio
\begin{equation}\label{eq:likelihood-primary-base}
  \mathbb{L}_\text{primary} =  \frac{p_g(\{\Delta_i,k_{t,i},z_i,\dots\})}{p_q(\{\Delta_i,k_{t,i},z_i,\dots\})}\, ,
\end{equation}
at single logarithmic accuracy in perturbative QCD, in the limit
where the emissions are strongly ordered in angle, i.e.\ that
$\Delta_1\gg \Delta_2\gg\dots\gg\Delta_n$.
In this limit, we should include in
Eq.~(\ref{eq:likelihood-primary-base}) the contributions associated
either with the running of the strong QCD coupling effects, or with
any hard-collinear effect.

The quark and gluon probability distributions can be computed
iteratively starting from the first (largest-angle) splitting.
A key point to take into account is the fact that collinear branchings
can change the flavour of the leading branch, either through a $g\to
q\bar q$ splitting, or through a $q\to qg$ splitting where the emitted
gluon is harder than the final quark.
At each splitting, we should therefore keep track of the flavour of the
leading parton as well as of its splitting fraction $z$ and its
relative transverse momentum $k_t$.
It is convenient to introduce a matrix
\begin{equation}\label{eq:probability-matrix}
  p_{ab}^{(i)}\equiv \begin{pmatrix}
    p(q_i|q_0) & p(q_i|g_0)\\
    p(g_i|q_0) & p(g_i|g_0)
  \end{pmatrix},
\end{equation}
where $p(b_i|a_0)$ denotes the probability that the harder branch has
flavour $b$ after the $i^\text{th}$ declustering, given that it
started (at step ``$0$'') with a jet of flavour $a$.
This matrix is initialised as $p_{ab}^{(0)}=\delta_{ab}$ and is
recursively constructed from step $i-1$ to step $i$ for each of the
$i=1,\dots,n$ Lund declusterings.

Assuming that just before branching $i$ the jet has flavour $a$, the
probability after branching $i$ should include two effects: (i) the
probability that the splitting has the observed kinematic properties
$\Delta_i, z_i, \dots$, potentially including a change of the leading
flavour, and (ii) a Sudakov factor implementing the fact that no
emission has occurred between the previous angle $\Delta_{i-1}$ and
$\Delta_i$ (with $\Delta_0\equiv R$), and with
$k_t>k_{t,\text{cut}}$. This Sudakov resums the virtual corrections
between $\Delta_{i-1}$ and $\Delta_i$.
This leads to the recursion
\begin{equation}\label{eq:probability-matrix-step}
  p_{ab}^{(i)}
  =\frac{\alpha_s(k_{ti})}{\pi \Delta_i}
  \begin{pmatrix}
    \tilde P_{qq}(z_i) & \tilde P_{qg}(z_i)\\
    \tilde P_{gq}(z_i) & \tilde P_{gg}(z_i)
  \end{pmatrix}
  \begin{pmatrix}
    S_q^{(i-1,1)} & 0\\
    0 & S_g^{(i-1,1)}
  \end{pmatrix}
  p_{ab}^{(i-1)}.
\end{equation}
In this expression, the splitting kernels $\tilde P_{ab}$ are directly
related to the Altarelli-Parisi splitting functions with the extra
requirement that since the declustering procedure follows the hardest
branch one should impose $z_i<\tfrac{1}{2}$:\footnote{We have chosen
  notations where the indices of the $\tilde P$ kernels represent the
  flavour of the hard branch, so as to make the matrix product
  in Eq.~\eqref{eq:probability-matrix-step} more obvious.
  As a consequence, these indices do not always match with the
  standard indices in the Altarelli-Parisi kernels where the indices
  instead refer to the flavour of the emitted parton with momentum
  fraction $z$.
  Finally, our probability distributions are taken differentially in
  $\Delta_i$ and $z_i$. The specific choice of variables is however
  irrelevant for the problem of quark-gluon classification as it
  cancels in the likelihood ratio.}
\begin{subequations}\label{eq:splitting-kernels}
  \begin{align}
    \tilde{P}_{qq}
    & = P_{gq}(z)\Theta(z<\tfrac{1}{2})
    = C_F \frac{1+(1-z)^2}{z}\Theta(z<\tfrac{1}{2}), \\
    \tilde{P}_{gq}
    & = P_{qq}(z)\Theta(z<\tfrac{1}{2})
    = C_F \frac{1+z^2}{1-z}\Theta(z<\tfrac{1}{2}), \\
    \tilde{P}_{qg}
    & = [P_{qg}(z)+P_{qg}(1-z)]\Theta(z<\tfrac{1}{2})
    = 2 n_f T_R [z^2+(1-z)^2]\Theta(z<\tfrac{1}{2}), \\
    \tilde{P}_{gg}
    & = [P_{gg}(z)+P_{gg}(1-z)]\Theta(z<\tfrac{1}{2})
    = 2C_A\left[\frac{1-z}{z}+\frac{z}{1-z}+z(1-z)\right]\Theta(z<\tfrac{1}{2}).
  \end{align}
\end{subequations}
The Sudakov factors, $S_{q,g}^{(i-1,i)}$, between the angle of the
last splitting $\Delta_{i-1}$ and the angle of the current splitting
$\Delta_i$ is computed as
\begin{equation}\label{eq:Sudakov}
  S_f^{(i-1,i)}
  = \exp\left[
    -\int_{\Delta_{i}}^{\Delta_{i-1}} \frac{d\Delta}{\Delta}\int dz
    \frac{\alpha_s(p_{ti}z\Delta)}{\pi}P_f(z_i)\Theta(p_{ti}z\Delta>k_{t,\text{cut}})        
  \right],
\end{equation}
with $P_f$ the total splitting function for a parton of flavour $f$
and $p_{ti}$ the transverse momentum (with respect to the beam) of
parton $i$ before splitting.
The $k_t$ of the emission is taken as $p_{ti}z\Delta$ which is
equivalent to our definition in Eq.~(\ref{eq:lund-coordinates}) in the
collinear limit.\footnote{Conversely, the value of $p_{ti}$ can be
  deduced from $\Delta_i$, $k_{ti}$ and $z_i$ using
  $p_{ti}=k_{ti}/(z_i\Delta_i)$.}
This Sudakov is evaluated at next-to-leading logarithmic (NLL)
accuracy with $\Delta_i\ll \Delta_{i-1}$, and we find
\begin{align}\label{eq:sudakov-nll}
  S_f^{(i-1,i)}=\exp\bigg\{&-\frac{C_f}{2\pi\alpha_s\beta_0^2}
   \bigg[
    (1-\lambda_{i-1}) \ln\frac{1-\lambda_{i-1}}{1-\lambda_\text{cut}}
    -(1-\lambda_{i}) \ln\frac{1-\lambda_{i}}{1-\lambda_\text{cut}}
    -\lambda_i+\lambda_{i-1}\nonumber\\
  & -\frac{\alpha_s\beta_1}{\beta_0}\left(
    \frac{1}{2}\ln^2(1-\lambda_i)-\frac{1}{2}\ln^2(1-\lambda_{i-1})
    +\frac{\lambda_i-\lambda_{i-1}}{1-\lambda_\text{cut}}\ln(1-\lambda_\text{cut})
    \right)\nonumber\\
  & +\left(\frac{\alpha_sK}{2\pi}-\frac{\alpha_s\beta_1}{\beta_0}\right)
    \left(\frac{\lambda_i-\lambda_{i-1}}{1-\lambda_\text{cut}}-\ln\frac{1-\lambda_{i-1}}{1-\lambda_i} \right)
  \bigg]\bigg\},
\end{align}
with $\alpha_s\equiv\alpha_s(p_{t,\text{jet}}R)$,
\begin{subequations}
\begin{align}
  \lambda_{i-1} &= 2\alpha_s\beta_0\left(\ln\frac{R}{x\Delta_{i-1}}-B_f\right),
  & \beta_0 & = \frac{11C_A-4n_fT_R}{12\pi},\\
  \lambda_i     &= 2\alpha_s\beta_0\left(\ln\frac{R}{x\Delta_i}-B_f\right),
  & \beta_1 & = \frac{17C_A^2-5C_An_f-3C_Fn_f}{24\pi^2},\\
  \lambda_\text{cut} &= 2\alpha_s\beta_0\ln\frac{p_{t,\text{jet}}R}{k_{t,\text{cut}}},
  & K & = \left(\frac{67}{18}-\frac{\pi^2}{6}\right)C_A - \frac{5}{9}n_f,\\
  B_q & = -\frac{3}{4}
  & B_g & = -\frac{11C_A-4n_fT_R}{12C_A},
\end{align}
\end{subequations}
and $x$ defined as the momentum fraction of the total jet momentum
carried by the subjet $j$ before branching $i$.
We note that in the rare occurrences where the Lund declusterings are
not ordered in angle --- which cannot be ruled out with the
Cambridge/Aachen declustering procedure --- we set the Sudakov to
$S_f=1$.
We also point out that the contribution from hard-collinear splitting
to the above expressions have been computed by setting an upper bound
$e^{B_f}$ on the $z$ integration in~(\ref{eq:Sudakov}). This is
correct at NLL accuracy. Although it has the drawback to insert
uncontrolled subleading corrections --- compared to the traditional
expression which can be recovered by keeping only the first
non-trivial term in $B_f$ --- it has the advantage of having a clean
endpoint, i.e.\ $S_f=1$ for $\lambda_{i-1}\le \lambda_\text{cut}$.

If we introduce the short-hand notations $\tilde{P}_{ab}^{(i)}$ and
$S_{ab}^{(i,i-1)}\equiv\delta_{ab}S_a^{(i,i-1)}$ for the full
splitting matrix and Sudakov matrix, the probabilities after including
all the Lund declusterings takes the form
\begin{equation}\label{eq:primary-collinear-full}
p^{\text{(final)}} = S^{(n+1,n)}\tilde{P}^{(n)}S^{(n,n-1)}\dots\tilde{P}^{(i)}S^{(i,i-1)}\dots\tilde{P}^{(1)}S^{(1,0)}p^{(0)},
\end{equation}
where the leftmost factor in the right-hand side takes into account
the fact that there are no more emissions between the angle of the
last declustering, $\Delta_n$, and the smallest angle accessible after
the last splitting:
$\Delta_{n+1}\equiv\Delta_\text{min}=k_{t,\text{cut}}/p_{tn}$ with
$p_{tn}$ the transverse momentum (with respect to the beam) of the
leading parton after the last declustering.
Eq.~\eqref{eq:primary-collinear-full} has the simple physical
interpretation of successive primary branchings, producing the factors
$\tilde P^{(i)}$, interleaved with Sudakov factors, $S^{(i,i-1)}$,
which resum virtual corrections between two primary emissions.
Finally, the probabilities associated with an initial quark or gluon
jet are given by
\begin{subequations}\label{eq:primary-probabilities-qg}
  \begin{align}
    p_q(\{\Delta_i,k_{t,i},z_i,\dots\})
    &= p^{\text{(final)}}(q|q_0) + p^{\text{(final)}}(g|q_0),\\
    p_g(\{\Delta_i,k_{t,i},z_i,\dots\})
    &= p^{\text{(final)}}(q|g_0) + p^{\text{(final)}}(g|g_0),
  \end{align}
\end{subequations}
translating the fact that we are inclusive over all flavours of
the final leading parton.

The probabilities in Eqs.~(\ref{eq:primary-probabilities-qg}) can be
directly inserted in~(\ref{eq:likelihood-primary-base}) to obtain a
quark-gluon discriminant.
It is, by construction, the optimal discriminant at single-logarithmic
accuracy in the limit where the declusterings are strongly ordered in
angle.
Since the above procedure keeps track of the flavour and momentum
fraction $x$ of the leading parton at each step, it takes into account
the possible correlations between the different declusterings, hence
going beyond the independent-emission assumption used with the average
Lund plane density (section~\ref{sec:average-primary-density},
Eq.~(\ref{eq:likelihood-density})).

%----------------------------------------------------------------------
\subsection{Extension to the full clustering tree}\label{sec:tree}

While non-primary (secondary, tertiary, ...) declusterings have no
impact at leading-logarithmic accuracy, they start carrying
information at our single-logarithmic accuracy.
Generalising the approach from the previous section to the full
clustering tree is mostly a technical step.
This time, we therefore settle to compute
\begin{equation}\label{eq:likelihood-tree-base}
  \mathbb{L}_\text{tree} =  \frac{p_g(\mathcal{L}_\text{tree})}{p_q(\mathcal{L}_\text{tree})}
\end{equation}
in the strongly angular-ordered limit.

This is again done recursively over the full (de-)clustering tree.
For this, consider a declustering $j_\text{parent}\to
j_\text{hard}+j_\text{soft}$, with kinematic variables $\mathcal{T}$,
i.e.\ with an angle $\Delta$, a soft momentum
fraction $z$ and relative transverse momentum $k_t$.
The probabilities associated with the parent jet can be deduced from
those of the subjets as follows:
\begin{subequations}\label{eq:tree-main-recursion}
  \begin{align}
    p_q(\mathcal{L}_\text{parent})
    & = S_q(\Delta_\text{prev},\Delta)\left[
      \tilde P_{qq}(z) p_q(\mathcal{L}_\text{hard}) p_g(\mathcal{L}_\text{soft}) + 
      \tilde P_{gq}(z) p_g(\mathcal{L}_\text{hard}) p_q(\mathcal{L}_\text{soft}) 
      \right ]\\
    p_g(\mathcal{L}_\text{parent})
    & = S_g(\Delta_\text{prev},\Delta)\left[
      \tilde P_{gg}(z) p_g(\mathcal{L}_\text{hard}) p_g(\mathcal{L}_\text{soft}) + 
      \tilde P_{qg}(z) p_q(\mathcal{L}_\text{hard}) p_q(\mathcal{L}_\text{soft}) 
      \right ]
  \end{align}
\end{subequations}
where $\Delta_\text{prev}$ is the angle at which the last
declustering before the one under consideration happened (with
$\Delta_\text{prev}=R$ for the first declustering). 
As in section~\ref{sec:collinear}, $S(\Delta_\text{prev},\Delta)$ is a
Sudakov factor imposing that no other emission with
$k_t>k_{t,\text{cut}}$ occurred since the last declustering at an
angle $\Delta_\text{prev}$, cf.\
Eq.~(\ref{eq:sudakov-nll}).\footnote{Although we have only made
  explicit the angular dependence, the Sudakov factors also depend on
  the prong momenta.}
The splitting kernels $\tilde P_{ab}$ are the same as in
Eq.~(\ref{eq:splitting-kernels}).
These expressions have the same form as
Eq.~(\ref{eq:probability-matrix-step}) except that, at each step, they
also include the probability for the soft branch.

This recursion is applied until each branch can no longer be
declustered in which case, if the last splitting has occurred at an
angle $\Delta_\text{last}$, one then just includes a factor 
\begin{subequations}\label{eq:tree-no-declustering}
  \begin{align}
    p_q(\mathcal{L}=\emptyset)
    & = S_q(\Delta_\text{last},\Delta_\text{min}),\\
    p_g(\mathcal{L}=\emptyset)
    & = S_g(\Delta_\text{last},\Delta_\text{min}),
  \end{align}
\end{subequations}
where, as for the primary case,
$\Delta_\text{min}=k_{t,\text{cut}}/p_t$ for a final branch of
momentum $p_t$.

%======================================================================
\section{Beyond strong angular ordering: Including clustering logarithms}\label{sec:clustering-logs}

%----------------------------------------------------------------------
\subsection{Generic considerations}\label{sec:clustering-logs-generic}

We conclude this section on analytic methods by discussing the
inclusion of clustering logarithms in our approach.
These logarithms arise from situations where we have at least two
emissions with commensurate angles and the exact Cambridge/Aachen
clustering has to be considered in order to label the emissions as
primary, secondary, ternary, etc.
When the emissions at commensurate angles are strongly ordered in
energy, this leads to single-logarithmic contributions (see
e.g.~\cite{Lifson:2020gua}).

In practice, the Cambridge/Aachen clustering can produce clusterings
which are not in agreement with the naive physical
expectation. Consider for example a quark-initiated jet with two
gluon emissions. The harder emission is emitted from the quark and
comes with a colour factor $C_F$. The softer emission can either be
seen as emitted from the quark, with a colour factor $C_F$, or as
emitted from the gluon, with a colour factor $C_A$. When the two
emission angles are similar, the actual Cambridge/Aachen clustering
will sometimes cluster the second gluon in the $C_F^2$ contribution
with the first gluon, yielding a secondary Lund declustering, or,
conversely, cluster the second gluon from the $C_FC_A$ contribution
with the hard quark, yielding a primary declustering.

In order to compute these contributions, we need the full angular
structure of the matrix elements for an arbitrary number of emissions
at commensurate angles. When computing the average Lund plane density,
this can be addressed, at least in the large-$N_c$ limit, by a Monte
Carlo integration similar to the one used to resum non-global
logarithms in~\cite{Dasgupta:2001sh}. (See~\cite{Hatta:2013iba} for an
approach valid at finite $N_c$.)

In our quark-gluon-discrimination application, one would have to keep
track of all the possible colour configuration with which an emission
can be radiated by the full set of emissions at larger transverse
momentum (or, at our accuracy, at larger energies). This is beyond
what can be practically achieved. Instead, we will adopt a simplified
approach where we apply a matrix-element correction which only
describes correctly situations where (any number of) pairs of
emissions are at commensurate angles.
This is similar in spirit to the NODS scheme introduced
in~\cite{Hamilton:2020rcu} to implement subleading-$N_c$ corrections
in parton showers.\footnote{Note the key difference that the NODS method 
  produces the correct behaviour at large-$N_c$ for any number of
  emissions at commensurate angles. The matrix-element correction only
  applies to subleading-$N_c$ corrections. In our case, the correct
  behaviour is only guaranteed for pairs of emissions at commensurate
  angles even in the large-$N_c$ limit.}

%------------------------------------------------------------------------
\subsection{Clustering logarithm with the full Lund tree}\label{sec:clustering-logs-full}

Since clustering logarithms have an explicit dependence on radiation
in/from different leaves, we first consider the situation where the
quark-gluon tagging is done using the full Lund declustering tree.
The case where only primary radiation is considered will be discussed
in section~\ref{sec:clustering-logs-prim} below.

In our approximation where we only allow for two emissions at
commensurate angles, we then consider two declusterings
$\mathcal{T}_1\equiv \{\Delta_1,k_{t1},z_1,\psi_1\}$ and
$\mathcal{T}_2\equiv \{\Delta_2,k_{t2},z_2,\psi_2\}$, with
$\Delta_1\sim\Delta_2\ll 1$. Since clustering corrections happen for
two emissions at similar angles and we only aim at describing the
configurations where we have only pairs of particles at commensurate
angles, we can assume that $\mathcal{T}_1$ and $\mathcal{T}_2$
correspond to consecutive Lund declusterings and that all the other
emissions are at widely different angles.
We can further assume that $\mathcal{T}_1$ happens before
$\mathcal{T}_2$ in the sequence of declusterings, i.e.\
$\Delta_1>\Delta_2$.
Our approach is to modify the emission probability in
Eq.~(\ref{eq:likelihood-tree-base}) for $\mathcal{T}_2$ to include
corrections due to the presence of $\mathcal{T}_1$.

\begin{figure}
  \centering
  \begin{subfigure}[t]{0.32\textwidth}
    \includegraphics[width=\textwidth,page=1]{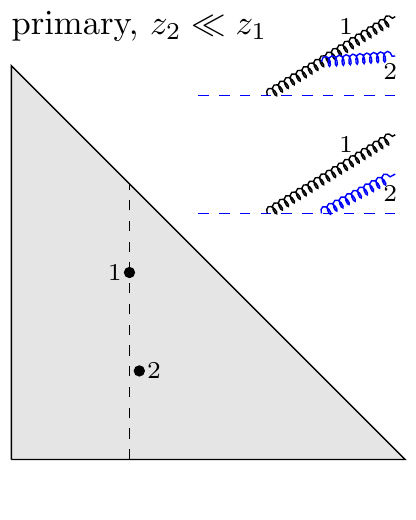}
    \caption{}\label{fig:clustering-illust-case1}
  \end{subfigure}
  \hfill
  \begin{subfigure}[t]{0.32\textwidth}
    \includegraphics[width=\textwidth,page=2]{figs/clustering-cases.pdf}
    \caption{}\label{fig:clustering-illust-case2}
  \end{subfigure}
  \hfill
  \begin{subfigure}[t]{0.32\textwidth}
    \includegraphics[width=\textwidth,page=3]{figs/clustering-cases.pdf}
    \caption{}\label{fig:clustering-illust-case3}
  \end{subfigure}
  \caption{Schematic representation of the three kinematic
    configurations affected by clustering logarithms. We consider two
    consecutive declusterings $\mathcal{T}_1$ and $\mathcal{T}_2$ with
    $\mathcal{T}_2$ occurring after $\mathcal{T}_1$ in the Lund
    sequence. If $\mathcal{T}_2$ is reconstructed as ``primary''
    (i.e. in the same plane as $\mathcal{T}_1$) it can either be much
    softer (case (a)) or much harder (case (b)) than
    $\mathcal{T}_1$. Case (c) describes the situation where
    $\mathcal{T}_2$ is reconstructed as a secondary emission from
    $\mathcal{T}_1$, and hence can be considered much
    softer.}\label{fig:clustering-illust}
\end{figure}

There are two main kinematic configurations to consider: either
$\mathcal{T}_1$ and $\mathcal{T}_2$ are both reconstructed as
consecutive ``primary'' emissions from the same hard branch, or
$\mathcal{T}_2$ is reconstructed as a ``secondary'' emission from
$\mathcal{T}_1$.\footnote{Primary and secondary are here counted from
  the hard branch common to both emissions, even if this one can be
  anywhere in the Lund tree.}
In the ``primary'' case, we can either have $z_2\ll z_1$ ($k_{t2}\ll
k_{t1}$) or $z_2\gg z_1$ ($k_{t2}\gg k_{t1}$), while in the ``secondary"
case we can assume $z_2\ll z_1$ ($k_{t2}\ll k_{t1}$).
This is illustrated by the Lund diagrams in
Fig.~\ref{fig:clustering-illust}.
At single-logarithmic accuracy, the clustering correction is computed
in the flavour channel where both emissions are gluons.
The distinction between the primary and secondary cases is decided by
the Cambridge/Aachen clustering. In both cases, if $C_R$ is the colour
factor of the common hard branch, the matrix element corresponding to
a given clustering sequence will have a contribution proportional to
$C_R^2$ and one proportional to $C_RC_A$. In the
strongly-angular-ordered limit, only the first term ($C_R^2$)
contributes to the ``primary'' clustering and only the second term
($C_RC_A$) to the ``secondary'' clustering.
The gluon-emission diagrams shown in Fig.~\ref{fig:clustering-illust}
represent the two contributions for each clustering case. The first
particles to cluster are highlighted in blue.

Let us first handle the case where $\mathcal{T}_1$ and $\mathcal{T}_2$
are both ``primary'' emissions. Say the parent parton has a colour
factor $C_R$. If $z_1\gg z_2$, the
$C_R\, {\rm d}^2\Delta_2/\Delta_2^2$ behaviour which corresponds to
the collinear limit in section~\ref{sec:tree} should be replaced by
the full soft-gluon radiation squared matrix element
\begin{equation}
  \left[
    \frac{C_A}{2}\frac{1}{\Delta_{12}^2} +
    \frac{C_A}{2}\frac{\Delta_1^2}{\Delta_{12}^2\Delta_2^2} +
    \left(C_R-\frac{C_A}{2}\right)\frac{1}{\Delta_2^2}
  \right] d^2\Delta_2,    
\end{equation}
where $\Delta_{12}^2 =
\Delta_1^2+\Delta_2^2-2\Delta_1\Delta_2\cos(\psi_2-\psi_1)$ is the
angle between the two emitted gluons.
This means that, in the gluon emission part of
Eq.~\eqref{eq:probability-matrix-step}, we should apply a correction
factor
\begin{equation}\label{eq:ommega-prim}
  \Omega_\text{prim} = 1 + \frac{C_A}{2C_R}\left(
    \frac{\Delta_2^2}{\Delta_{12}^2} + 
    \frac{\Delta_1^2}{\Delta_{12}^2} - 1
  \right).
\end{equation}
It is straightforward to show that the ``primary'' case with $z_1\ll
z_2$,  gives the same correction $\Omega_\text{prim}$.
As expected, $\Omega_\text{prim}\to 1$ when $\Delta_1\gg\Delta_2$ (or
when $\Delta_1\ll\Delta_2$) so that the strongly-ordered limit is
recovered. Since both emissions are primary, we never have $\Delta_{12}\ll\Delta_1,\Delta_2$.

\begin{figure}
  \centering
  \begin{subfigure}[t]{0.48\textwidth}
    \includegraphics[width=\textwidth,page=1]{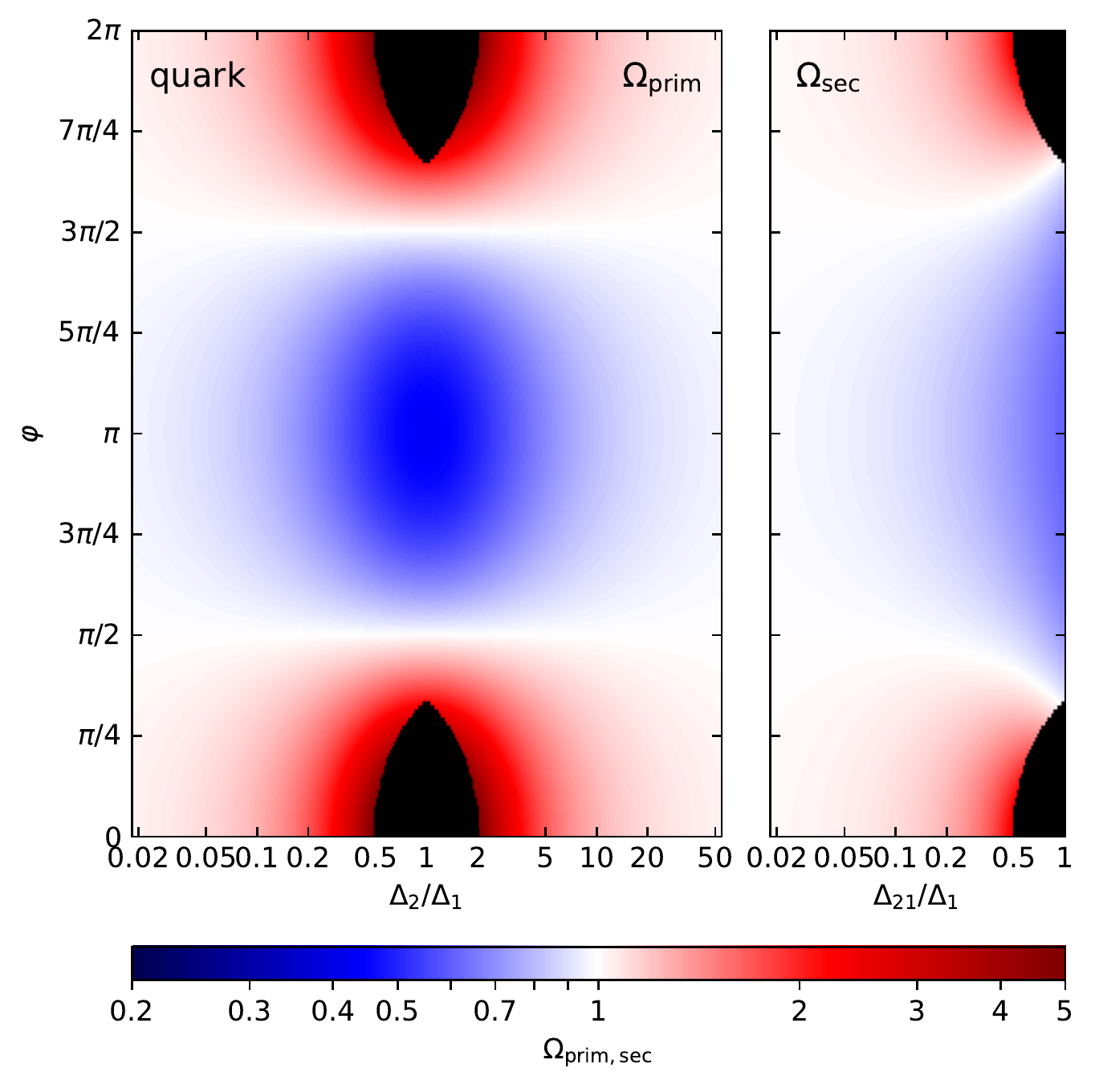}
    \caption{}
  \end{subfigure}
  \hfill
  \begin{subfigure}[t]{0.48\textwidth}
    \includegraphics[width=\textwidth,page=2]{figs/plot-omega.pdf}
    \caption{}
  \end{subfigure}
  \caption{Clustering correction factors $\Omega_\text{prim}$
    (primary) and $\Omega_\text{sec}$ (secondary) for quark (left) and
    gluon (right) leading partons. The black regions correspond to the
    kinematic regions where emissions are not clustered in the
    indicated Lund plane.}\label{fig:clustering-corr}
\end{figure}

We now turn to the ``secondary'' case where $\mathcal{T}_2$ is emitted
from the soft branch of $\mathcal{T}_1$.
Here, $\Delta_2$ is the angle between the emissions $\mathcal{T}_1$
and $\mathcal{T}_2$, i.e.\ $\Delta_2\equiv\Delta_{12}$, and we denote by
$\Delta_{02}=
\Delta_1^2+\Delta_2^2-2\Delta_1\Delta_2\cos(\psi_2-\psi_1)$ the angle
between $\mathcal{T}_2$ and the hard branch of $\mathcal{T}_1$. In the
collinear limit this would correspond to a factor $C_A/\Delta_2^2$
which has to be replaced by the full angular structure
\begin{equation}
    \frac{C_A}{2}\frac{1}{\Delta_2^2} +
    \frac{C_A}{2}\frac{\Delta_1^2}{\Delta_{02}^2\Delta_2^2} +
    \left(C_R-\frac{C_A}{2}\right)\frac{1}{\Delta_{02}^2},
\end{equation}
yielding a correction factor
\begin{equation}\label{eq:ommega-sec}
  \Omega_\text{sec} = \frac{1}{2}\left(
    1 + \frac{\Delta_1^2}{\Delta_{02}^2} + 
    \frac{\Delta_2^2}{\Delta_{02}^2}\right) +
  \frac{C_R}{C_A}\frac{\Delta_2^2}{\Delta_{02}^2}.
\end{equation}
Without surprise, $\Omega_\text{sec}\to 1$ if $\Delta_2\ll\Delta_1$,
recovering the strongly-ordered case.
The correction factors $\Omega_\text{prim}$ and $\Omega_\text{sec}$
are plotted in Fig.~\ref{fig:clustering-corr}. We see that they are
indeed localised around the region where both emissions have
commensurate angles. They tend to be larger for quarks than for
gluons.

It is interesting to notice that the above correction which accounts
for clustering logarithms introduces a dependence om the azimuthal
angle $\psi$. It is the only dependence on $\psi$ at the single
logarithmic accuracy.

In principle, the Sudakov factors should also receive
single-logarithmic corrections due to clustering effects. Since
clustering logarithms only affect flavour-diagonal emissions of two
soft gluons (at least within our approximations), it is however
relatively straightforward to convince oneself that these
corrections only lead to a reshuffling of some contributions
between different factors in the overall probability distribution and
can therefore be neglected.
This can be understood as follows. 
Suppose one has a parton emitting a gluon ``1'' at an angle
$\Delta_1$. In our approach, that the previous emission occurred at an
angle $\Delta_0$ and that the following emissions on the hard and soft
branches occur at angles $\Delta_{2,\text{hard}}$ and
$\Delta_{2,\text{soft}}$ respectively.
In the strongly-ordered limit, emission ``1'' is involved in three
contributions to the Sudakov: one between $\Delta_1$ and $\Delta_0$,
one between $\Delta_{2,\text{hard}}$ and $\Delta_1$, and one between
$\Delta_{2,\text{soft}}$ and $\Delta_1$.
Within our approximation, if we want to compute the corrections to the
Sudakov at angles commensurate with $\Delta_1$, we can assume that all
the other emissions are at widely separate angles, i.e.\
$\Delta_{2,\text{hard}},\Delta_{2,\text{soft}}\ll \Delta_1\ll\Delta_0$.
To compute the overall Sudakov factor, summed over the three regions
described above, one should integrate the exact matrix
element, including the full angular structure.
Within our approximation where we only target correctness for two
emissions at commensurate angles, this integral is proportional to
$C_R\log(\Delta_0/\Delta_{2,\text{hard}})+C_A\log(\Delta_1/\Delta_{2,\text{soft}})$,
which is the same as the strongly-ordered limit.

In practice, inserting corrections factors due to clustering
logarithms in Eq.~(\ref{eq:tree-main-recursion}) requires some care as
it depends whether the Lund declustering, $\mathcal{T}$, that it
implements comes from following the hard or the soft branch at the
previous declustering.
First, at our single-logarithmic accuracy, clustering corrections are
only non-trivial for two successive gluon emissions.
Then, say that the previous branching, happening at an angle
$\Delta_\text{prev}$ is denoted by $\mathcal{T}_\text{prev}$.
If $\mathcal{T}$ follows $\mathcal{T}_\text{prev}$ along the harder
branch, we only apply a correction for the contributions where $\mathcal{T}_\text{prev}$ did
not have a flavour change. The correction is then applied only for the
flavour-diagonal contribution with a colour factor $C_R$ being $C_F$
or $C_A$ depending on the flavour of the hard parton.
Conversely, if $\mathcal{T}$ follows $\mathcal{T}_\text{prev}$ along
the softer branch, only the flavour-diagonal term in
$p_g(\mathcal{L}_\text{parent})$ receives a correction with a colour
factor $C_R$ given by the flavour of the hard branch at the branching
$\mathcal{T}_\text{prev}$.

It is interesting to note that the correction factors
$\Omega_\text{prim}$ and $\Omega_\text{sec}$ explicitly depend on the
azimuthal angles of the declusterings, which is new compared to the
strongly-angular-ordered case.
If we want to consider only the $\Delta$, $z$ and $k_t$ variables for
each Lund declusterings (see the discussion in
section~\ref{sec:mc-full} below), we can integrate out the $\phi$
dependence, averaging over the domain allowed by the fact that the
declustering $\mathcal{T}_1$ is undone before $\mathcal{T}_2$. We
denote these azimuthally-averaged correction factors by
$\bar\Omega_\text{prim}$ and $\bar\Omega_\text{sec}$. They only depend
on the ratio $x=\Delta_2/\Delta_1$ and are found to be
\begin{subequations}\label{eq:omega-average}
\begin{align}
  \bar\Omega_\text{prim}
  & \overset{x<1/2}{=} 1+\frac{C_A}{C_R}\frac{x^2}{1-x^2} \\
  & \overset{x>1/2}{=} 1+\frac{C_A}{2C_R}\left[\frac{1+x^2}{|1-x^2|}
    \frac{1-\frac{2}{\pi}\tan^{-1}\left(
    \frac{1+x}{|1-x|}\sqrt{\frac{2x-1}{2x+1}}
    \right)}{1-\frac{2}{\pi}\tan^{-1}\left(
    \sqrt{\frac{2x-1}{2x+1}}
    \right)}\right]\\
  \bar\Omega_\text{sec}
  & \overset{x<1/2}{=} 1+\frac{C_R}{C_A}\frac{x^2}{1-x^2} \\
  & \overset{x>1/2}{=} \frac{1}{2}+
    \left[\frac{\text{sgn}(1-x^2)}{2}+\frac{C_R}{C_A}\frac{x^2}{|1-x^2|}\right]
    \frac{1-\frac{2}{\pi}\tan^{-1}\left(
    \frac{1+x}{|1-x|}\sqrt{\frac{2x-1}{2x+1}}
    \right)}{1-\frac{2}{\pi}\tan^{-1}\left(
    \sqrt{\frac{2x-1}{2x+1}}
    \right)}.
\end{align}
\end{subequations}

%------------------------------------------------------------------------
\subsection{Clustering logarithms with primary radiation
  only}\label{sec:clustering-logs-prim}

Our last analytic step is to include the effect of clustering
logarithms in the Lund quark-gluon discriminant which only uses
primary declusterings.
We do this in an approximation where we only allow for pairs of
emissions to be at commensurate angles.

As for the full Lund tree, two types of corrections should be
included: corrections to the matrix element for the radiation of two
soft gluons at commensurate angles, and potential corrections to the
Sudakov factor.
Corrections to the real radiation are trivial: the splitting factors
$\tilde P_{qq}(z_i)$ and $\tilde P_{gg}(z_i)$
in Eq.~\eqref{eq:probability-matrix-step} should be multiplied by a factor
$\Omega_\text{prim}$ (Eq.~(\ref{eq:ommega-prim})), respectively with
$C_R=C_F$ and $C_R=C_A$.

Since we are no longer including a Sudakov factor for the soft branch,
our previous argument saying that the overall Sudakov factor was not
affected by clustering effects no longer holds.
Let us therefore again consider a gluon emission ``1'' at an angle
$\Delta_1$ and relative transverse momentum $k_{t1}$.
In the soft-gluon limit, the total Sudakov factor at (relative)
transverse momentum scales smaller than $k_{t1}$ should use the full
matrix element for radiation from the system including both the parent
parton and emission ``1'', i.e.
\begin{align}
  -\log S_f^{(i-1,i)} =
  \int_{k_{t,\text{cut}}}^{k_{t1}}\frac{\alpha_s(k_t)}{\pi^2}
  \int d^2\Delta_2
  & \left[\left(C_R-\frac{C_A}{2}\right)\frac{1}{\Delta_2^2}
    +\frac{C_A}{2}\frac{1}{\Delta_{12}^2}+\frac{C_A}{2}\frac{\Delta_1^2}{\Delta_2^2\Delta_{12}^2}\right]\,\nonumber\\
  & \left[1-\Theta(\Delta_{12}<\Delta_1)\Theta(\Delta_{12}>\Delta_2)\right],
\end{align}
where the square bracket in the second line imposes that ``2'' is
clustered as a primary emission.
This gives a correction compared to the strongly-angular-ordered case
which is found to be\footnote{This result is essentially the same as the
  $\mathcal{O}(\alpha_s^2)$ clustering logarithms contribution to the
  primary Lund plane density found in~\cite{Lifson:2020gua} (see
  Eq.~(3.25) there).}
\begin{equation}\label{eq:clustering-sudakov-prim}
  -\delta \log S_f^{(i-1,i)} = (C_A-C_R)\xi
  \left[\int_{k_{t,\text{cut}}}^{k_{t1}}\frac{\alpha_s(k_t)}{\pi}\right],
\end{equation}
with $\xi=0.323006$.
Note that this contribution happens to vanish when the parent parton
is a gluon ($C_R=C_A$).

%======================================================================
\section{Machine learning approaches}\label{sec:ml-approaches}

%----------------------------------------------------------------------
\subsection{Primary Lund plane and LSTM}\label{sec:lstm}

A natural approach to adopt is the Deep-Learning technique used in the
original study of the primary Lund Plane~\cite{Dreyer:2018nbf}, which
was already showing an excellent discriminating power in the context
of boosted $W$ tagging.
Here we only consider the long short-term memory (LSTM)~\cite{lstm1997} network
architecture as it showed the best performance
in~\cite{Dreyer:2018nbf}.

In practice, we input the list of Lund declusterings
$\{(\ln\Delta_i,\ln k_{ti})\}$ to an LSTM of dimension 128 connected to a
dropout layer (with rate 20\%), with a final dense layer of dimension two
and softmax activation.
The network weights are initialised with a He uniform variance scaling
initialiser~\cite{DBLP:journals/corr/HeZR015}, and the training is
performed using an Adam optimisation
algorithm~\cite{DBLP:journals/corr/KingmaB14} with a batch size of
128, a learning rate of 0.0005 and a categorical cross-entropy loss
function.
Our model is implemented using \texttt{TensorFlow}~v2.1.0.

The data sample is split into a training sample (80\%), a validation
sample (10\%) and a testing sample (10\%).
We train over a maximum of 50 epochs, with an early stopping when the
performance does not increase over four epochs.

For each configuration, we have run five independent trainings.
For the quality measures reported below, the central value is
obtained by averaging over the five runs and the uncertainty band is
taken as their envelope.

%----------------------------------------------------------------------
\subsection{Full Lund tree and Lund-Net}\label{sec:graph}

In order to take full advantage of the information contained in
secondary leaves of the Lund plane, we consider the Lund-Net model
introduced in Ref.~\cite{Dreyer:2020brq} and its associated
code~\cite{lundnet_code}.

As input, we transform the tree of Lund declusterings into a graph,
with the kinematic variables $\mathcal{T}$ of a declustering serving
as attributes of a node on the graph.
The Cambridge/Aachen clustering sequence is used to form bidirectional
edges along the nodes connected in the Lund tree.

The graph architecture uses an EdgeConv operation~\cite{DGCNN}, which applies a
multi-layer perceptron (MLP) to each incoming edge of a node, using combined
features of the node pairs as inputs, producing a learned edge
feature.
This initial shared MLP consists of two layers, each consisting of a
dense network with batch
normalisation~\cite{DBLP:journals/corr/IoffeS15} and ReLU
activation~\cite{glorot2011deep}, which are followed by an aggregation
step taking an element-wise average of the learned edge features of
the incoming edges as well as a shortcut connection~\cite{he2016deep}.
The same MLP is applied to all nodes, leading to updated node features
but keeping the structure of the graph unchanged.
The Lund-Net architecture consists of six such EdgeConv blocks stacked
together, and the number of channels of the MLPs are $(32, 32)$,
$(32, 32)$, $(64, 64)$, $(64, 64)$, $(128, 128)$ and $(128, 128)$.
Their output is concatenated for each node, and processed by a MLP
with 384 channels, to which a global average pooling is applied to
extract information from all nodes in the graph.
This is followed by a final fully connected layer with 256 units and a
dropout layer with rate 10\%, with a softmax output giving the result
of the classification.
The Lund-Net model is implemented with the Deep Graph Library
0.5.3~\cite{wang2020deep} using the PyTorch~1.7.1
\cite{NEURIPS2019_9015} backend, and training is performed for 30
epochs, using an Adam optimiser~\cite{DBLP:journals/corr/KingmaB14} to
minimise the cross entropy loss. An initial learning rate of 0.001 is
used, which is lowered by a factor 10 after the 10th and 20th epochs.
As for the LSTM approach, the data sample is randomly split in
80/10/10\% training/validation/testing samples, and we take the
average and envelope of five runs.

In this paper, the inputs for each Lund declustering include, by
default, $\ln\Delta$, $\ln k_t$, $\ln z$ and $\psi$.
In section~\ref{sec:mc-collinear} which probes the collinear limit of
our discriminants, the azimuthal angle is irrelevant and therefore not
included in any of our approaches.
Furthermore, in section~\ref{sec:v-others} we discuss the effect of
adding particle-ID information to the inputs, and in
section~\ref{sec:clust-azimuth-effects} we discuss the effect of the
azimuthal angle $\psi$.
When imposing a cut on the (relative) transverse momentum, only the
Lund declusterings with $k_t$ above the cut are included in the data
sample.

%======================================================================
\section{Validation in a pure-collinear (toy) parton shower}\label{sec:mc-collinear}

Before turning to a full Monte Carlo-based assessment of the
discriminating performance of the tools introduced in the previous
sections, we provide a cross-validation between the analytic 
and deep-learning approaches.
To do this, we use a setup in which our analytic approach in
sections~\ref{sec:collinear} and~\ref{sec:tree} corresponds to the
exact likelihood-ratio discriminant.
This is achieved by generating events directly in the
strong-angular-ordered limit, where our analytic approach from
section~\ref{sec:discrim-analytic} becomes exact:

For simplicity, we use a fixed-coupling approximation, a fixed initial
jet $p_t$ of 1~TeV (with $R$=1) and a fixed cut on emissions
$k_{t,\text{cut}}=1$~GeV. We simulate branchings using the full
Altarelli-Parisi splitting functions, keeping track at each emission
of the angle $\Delta$ and energy fraction $z$ of the emission.
In the strict collinear limit, a parton of momentum $p_t$ branches in
two partons of momenta $(1-z)p_t$ an $zp_t$, so the transverse
momentum of each parton in the cascade --- or, equivalently, its
fraction of the initial jet $p_t$ --- can be deduced from the angles
($\Delta_i$) and momentum fractions ($z_i$) at each branching.
In practice, we have used (a slightly adapted version of)\footnote{Our
  adaptation compared to the original work in~\cite{Dasgupta:2014yra}
  mostly consists in imposing a $k_t$ cutoff (instead of a small cut
  on $z$, as well as to keep the full tree of the generated cascade
  rather than just the final particles.}  the {\tt microjet}
code~\cite{Dasgupta:2014yra} to simulate events with strong angular
ordering.
The Lund declusterings are taken directly from the event trees,
without any reclustering with the Cambridge/Aachen jet algorithm. This
guarantees the absence of clustering logarithms.

Our analytic approaches to quark-gluon discrimination are applied as
described in sections~\ref{sec:collinear} and~\ref{sec:tree} except
for two details: (i) they have been adapted to use a fixed-coupling
approximation, and (ii) the Sudakov factor in Eq.~(\ref{eq:Sudakov})
has been computed keeping the full splitting function so as to
guarantee that the resulting probability distributions match exactly
that of the generated sample, including corrections strictly beyond
our single-logarithmic approximation. With a fixed-coupling
approximation, the calculation of the Sudakov exponent is relatively
straightforward and expressions are given in
appendix~\ref{sec:sudakov-full-splitting} for completeness.

\begin{figure}
  \centering
  \includegraphics[width=0.48\textwidth,page=1]{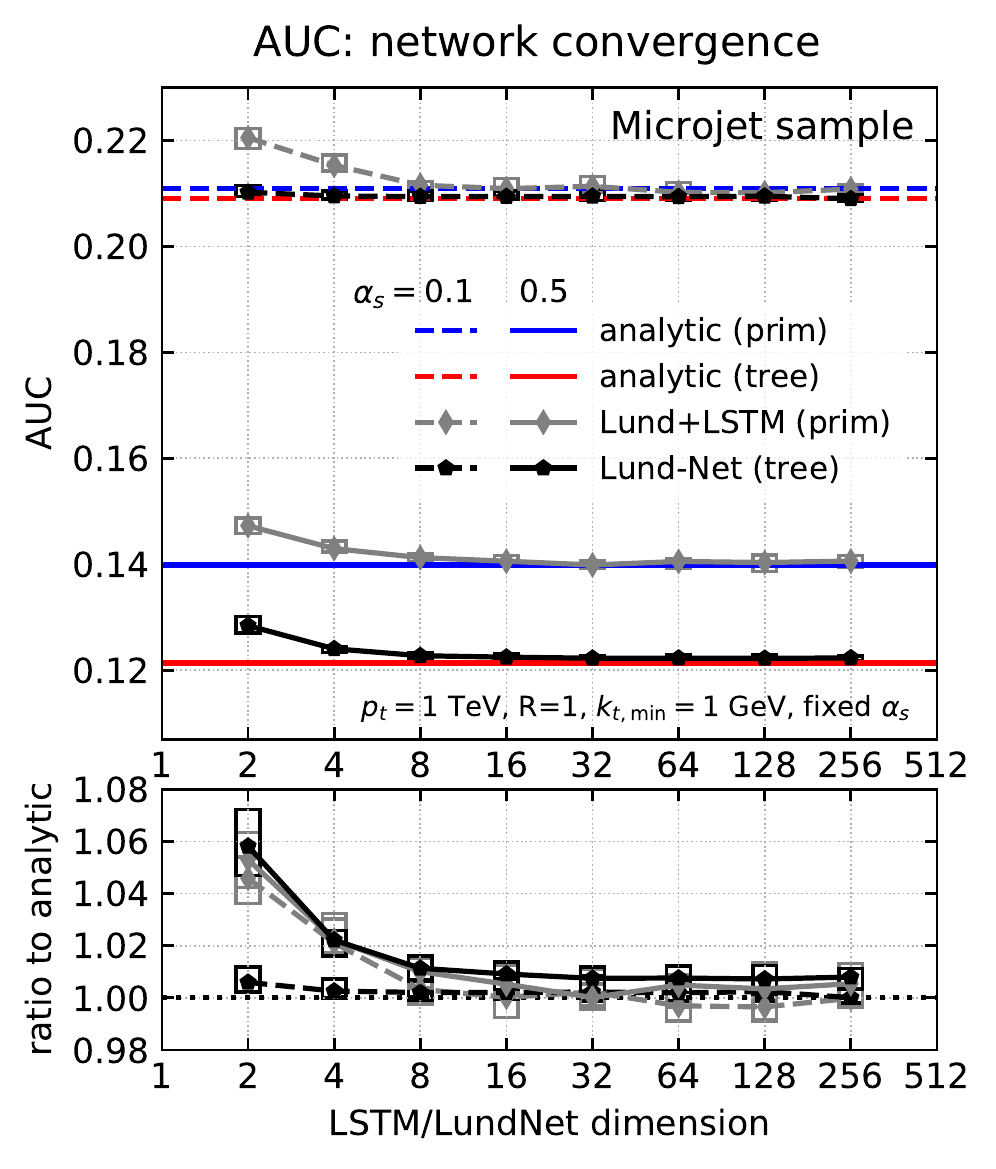}
  \caption{Area under the ROC curve (AUC) as a function of dimension
    of the network. The networks converge to the optimal (analytic)
    discriminant above $\approx 32$ nodes, proving that the networks
    learned all the features of the {\tt microjet} samples.
  }\label{fig:collinear-auc}
\end{figure}

With this setup in mind, we want to check that the machine learning
(ML) approach using an LSTM network trained on primary Lund
declusterings (section~\ref{sec:lstm}) converges to the same
performance as what is given by the analytic approach in
section~\ref{sec:collinear}.
Similarly, we expect that the Lund-Net approach from
section~\ref{sec:graph}, trained on full Lund trees, gives the same
performance as that of the analytic discriminant based on the full
Lund tree in section~\ref{sec:tree}.
We also want to check that these new tools offer a better
discriminating power than what is obtained using either the Iterated
Soft Drop multiplicity or the average primary Lund plane density (see
section~\ref{sec:average-primary-density}).

In practice, we use a sample of $10^6$ events generated with our
adapted version of the {\tt microjet} code, with $\alpha_s$ fixed 
either to 0.1 or to 0.5.
These samples are either used to compute the analytic
discriminant,\footnote{For $\alpha_s=0.1$, our analytic results have
  been obtained with a sample size of $10^7$ events instead of the
  default sample of $10^6$ events. This shows no visually observable
  differences on the results presented here.}
Eqs.~(\ref{eq:likelihood-primary-base})
or~(\ref{eq:likelihood-tree-base}) or as inputs to train/validate/test
our neural-network-based models.
For the methods using machine-learning, the event sample is split in
80/10/10\% training/validation/testing samples.
This is repeated five times for 5 different subdivisions and we take
the average and envelope of these five runs respectively as an
estimate of the performance and of the associated uncertainty.

\begin{figure}
  \centering
  \begin{subfigure}[t]{0.48\textwidth}
    \includegraphics[width=\textwidth,page=1]{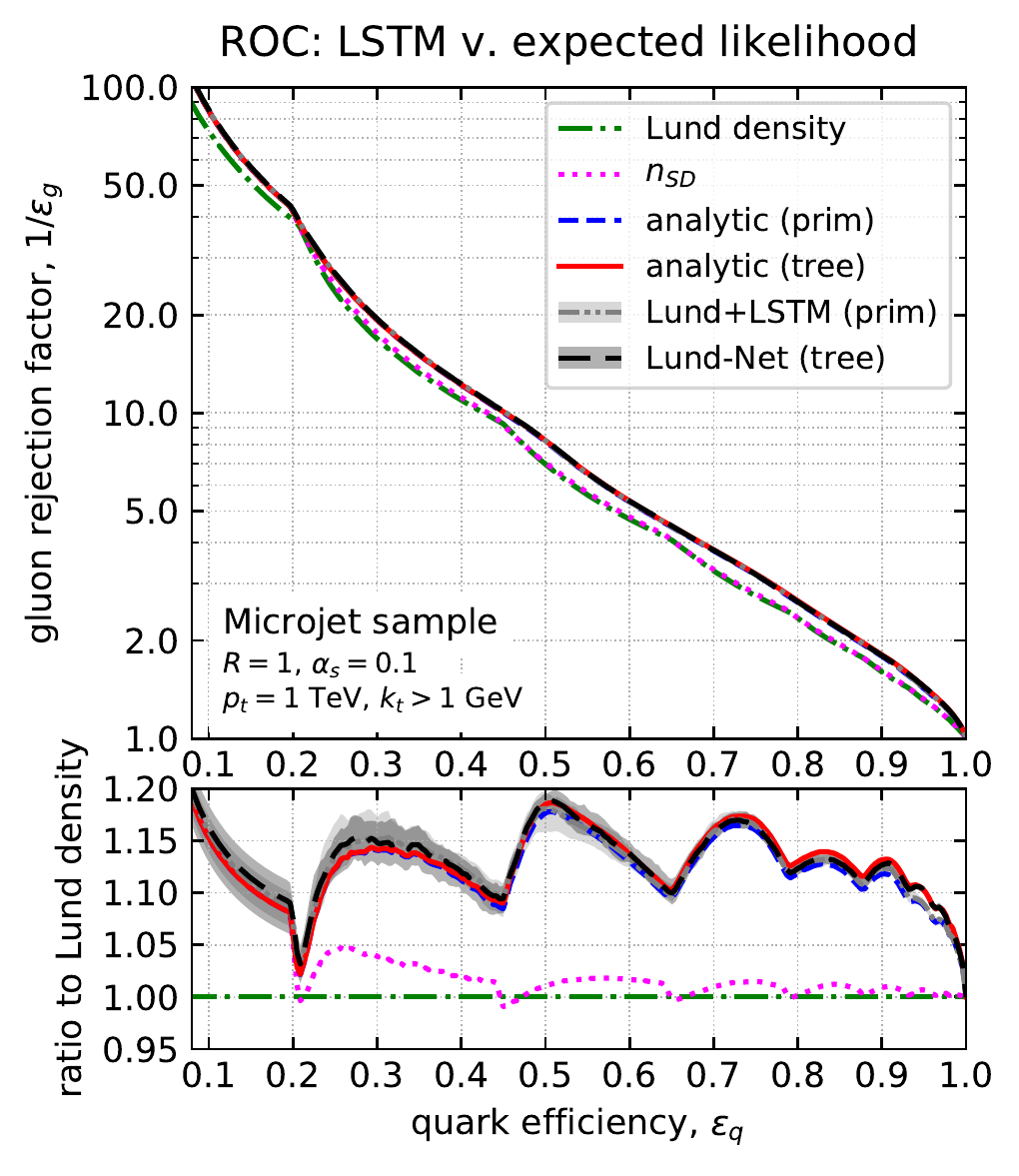}
    \caption{$\alpha_s=0.1$}\label{fig:collinear-roc-01}
  \end{subfigure}
  \hfill
  \begin{subfigure}[t]{0.48\textwidth}
    \includegraphics[width=\textwidth,page=2]{figs/plot-collinear-roc.pdf}
    \caption{$\alpha_s=0.5$}\label{fig:collinear-roc-05}
  \end{subfigure}
  \caption{The ROC curves for the analytic and ML classifiers for the
    {\tt microjet} sample. For this sample, our analytic approach is
    exact, showing that the neural networks capture the full
    (single-logarithmic) information in the training
    sample.}\label{fig:collinear-roc}
\end{figure}

Since the analytic models reproduce the exact likelihood ratio for
these collinear samples, they are expected to provide the optimal
discriminants.
We first study how the area under the ROC curve (AUC) evolves as a
function of the dimension of the LSTM or EdgeConv block in our
machine-learning setup, varied between 2 and 256 nodes,%
\footnote{For Lund-Net, the dimension refers to the size of the first MLP
  in the initial EdgeConv block, keeping the scaling of the successive layers
  identical to the one in section~\ref{sec:graph}.}
compared to the expected exact analytic result.
This is shown in Fig.~\ref{fig:collinear-auc}.
It is remarkable that for a network dimension of 32 or above, the neural
network is able to reproduce the expected optimal discriminant to within
at most 1\%, for both values of $\alpha_s$.
If we look at the full ROC curves, Fig.~\ref{fig:collinear-roc}, we see
again the same level of agreement. The larger uncertainty at
smaller quark efficiencies is expectable as only a fraction of the
background events pass the tagger.
Based on Figs.~\ref{fig:collinear-auc} and ~\ref{fig:collinear-roc},
we note a hierarchy between the classifiers with Lund-based methods
performing better than the Iterated SoftDrop multiplicity and, among
the Lund-based methods, the ones using the full tree information
performing better that the ones using only primary declusterings.
Improving the (logarithmic) accuracy of the analytic approach and
exploiting more jet substructure information both lead to a
performance increase.
Furthermore, the performance differences are enhanced as one opens up
the phase space to include more emissions (or increase $\alpha_s$).

\begin{figure}
  \centering
  \begin{subfigure}[t]{0.48\textwidth}
    \includegraphics[width=\textwidth,page=2]{figs/plot-collinear-v-sample-size.pdf}
    \caption{AUC}\label{fig:collinear-v-sample-size-auc}
  \end{subfigure}
  \hfill
  \begin{subfigure}[t]{0.475\textwidth}
    \includegraphics[width=\textwidth,page=1]{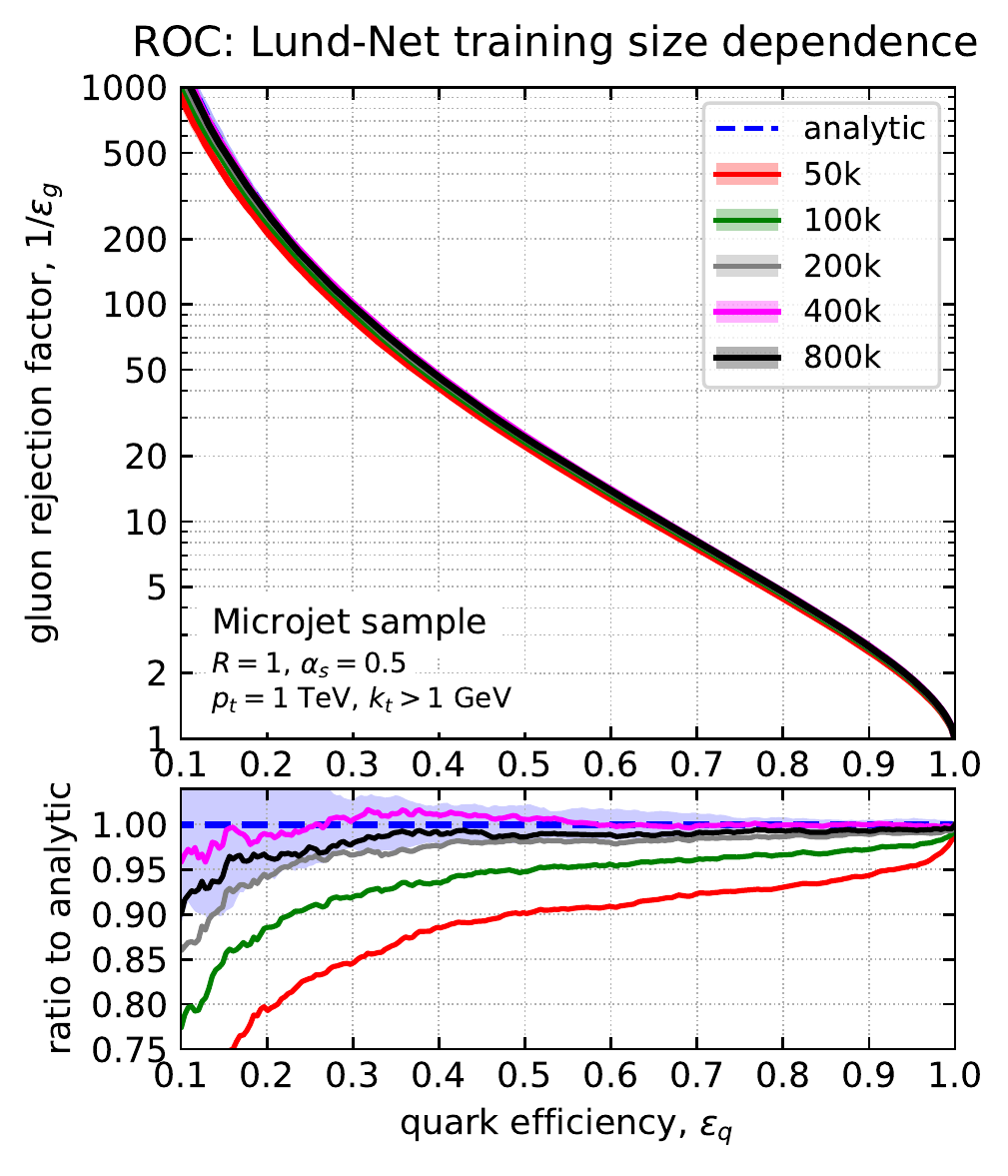}
    \caption{ROC curves}\label{fig:collinear-v-sample-size-roc}
  \end{subfigure}
  \caption{Convergence of the ML-based Lund taggers as a function of
    the training sample size for the AUC (a) and ROC curve (b). The
    ML testing phase is always performed on the same 100k events and the
    shaded band around the analytic results is the fluctuations across
    different samples of 100k events.
    The optimal performance is reached for training sample sizes of
    400k events and above.}\label{fig:collinear-v-sample-size}
\end{figure}

To further investigate the $\lesssim 1\%$ difference between the
Lund-Net and analytic results for $\alpha_s=0.5$, we show in
Fig.~\ref{fig:collinear-v-sample-size} the performance this time as a
function of the size of the training sample (keeping the size of the
validation and testing samples to $10^5$ jets).
The shaded band around the analytic expectation represent the
statistical fluctuations obtained by splitting the full $10^6$ event
sample in 10 subsamples of $10^5$ events, running our analytic
discriminant independently on  each subsample. For a testing sample
size of $10^5$ jets and $\varepsilon_q=0.1$, we only keep $\sim
0.1\%$, i.e.\ $\sim 100$, 
of the gluon jets which is compatible with the $\sim 10\%$ observed
statistical uncertainty.
One sees that within the statistical uncertainties, the performance of
the Lund-Net approach matches that of the analytic expectation for a
training sample of $4\times 10^5$ events or more. This is seen both
for the AUC, Fig.~\ref{fig:collinear-v-sample-size-auc} and for the
ROC curves, Fig.~\ref{fig:collinear-v-sample-size-roc}. In the latter
case, the convergence is slightly slower at small $\varepsilon_q=0.1$,
as one could have expected.

Before we close this section, we note that additional tests targetting
the asymptotic single-logarithmic limit of full Monte Carlo
simulations will be carried on in section~\ref{sec:asymptotic-tests}.

%======================================================================
\section{Full Monte Carlo simulations}\label{sec:mc-full}

\subsection{Setup}\label{sec:mc-setup}

We now move to testing the performance of Lund-plane-based quark-gluon
discriminants with full Monte Carlo samples.
For our reference quark and gluon samples, we simulated respectively
$Z+q$ and $Z+g$ events with Pythia
v8.24~\cite{Sjostrand:2006za,Sjostrand:2014zea} with multi-parton
interactions enabled with the Monash13 tune~\cite{Skands:2014pea}.
For the $Z+q$ sample, only light quark flavours ($u$, $d$ and $s$)
have been included. The $Z$ boson is set to decay to invisible
neutrinos.
Jets are then reconstructed on the remaining final-state using the
anti-$k_t$ algorithm~\cite{Cacciari:2008gp} with $R=0.4$, as
implemented in FastJet~\cite{Cacciari:2005hq,Cacciari:2011ma}.
We select at most the two hardest jets within $|y|<2.5$ and keep only
the ones with $500<p_t<550$~GeV. For each selected jet, we recluster
its constituents with the Cambridge/Aachen algorithm and we construct
the Lund declusterings following the recipe described in
section~\ref{sec:earlier}.
The studies described below are performed with quarks and gluon
samples of $10^6$ jets each. 

To probe the resilience of our quark-gluon discriminants against
various effects we have generated additional event samples.
The first one uses the same setup as above with hadronisation and
multi-parton interactions switched off, hence probing the influence of
non-perturbative effects.
The second uses dijet events, $qq\to qq$ with light quark flavours and
$gg\to gg$, and is meant to probe the dependence on the hard process.
The third one uses the same setup as the reference sample ($Z$+jet
with non-perturbative effects enabled), this time
generated with Herwig v7.2.0~\cite{Bahr:2008pv,Bellm:2015jjp} so as
to probe the dependence on the Monte Carlo generator.

We test a total of six quark-gluon discriminants: the Iterated Soft
Drop multiplicity ($n_\text{SD}$) and the discriminant based on the
average Lund plane density (Lund density), both described in
section~\ref{sec:earlier}, our new analytic discriminants using either
the primary declusterings only (analytic(prim),
sections~\ref{sec:collinear} and~\ref{sec:clustering-logs-prim}) or
the full declustering tree (analytic(tree), sections~\ref{sec:tree}
and~\ref{sec:clustering-logs-full}), and the deep-learning approaches
using either only the primary Lund declusterings (Lund+LSTM (prim)) or
the full Lund tree (Lund-Net (tree)) both described in
section~\ref{sec:ml-approaches}.

For the analytic models, clustering contributions are included with
their dependence on the azimuthal angle $\psi$. We further discuss the
influence of clustering logarithms and of the azimuthal angle $\psi$
in section~\ref{sec:clust-azimuth-effects} below.
Our analytic models are only considered in the presence of a $k_t$ cut
on the Lund declusterings, guaranteeing infrared-and-collinear
safety.

As in the previous section, for the methods using machine learning,
the event sample is subdivided into $8\times 10^5$ training jets,
$10^5$ evaluation jets and $10^5$ testing jets. We use five
different subdivisions of the full sample to assess the uncertainties
on the performance.
For the discriminant based on the average Lund plane density, we use
the first $9\times 10^5$ events to build a (binned) estimate of
$\rho_{q.g}(\Delta,k_t)$ and the $10^5$ remaining events as a testing
sample.

\subsection{Tagging performance}\label{sec:mc-performance}

\begin{figure}
  \centering
  \includegraphics[width=0.48\textwidth]{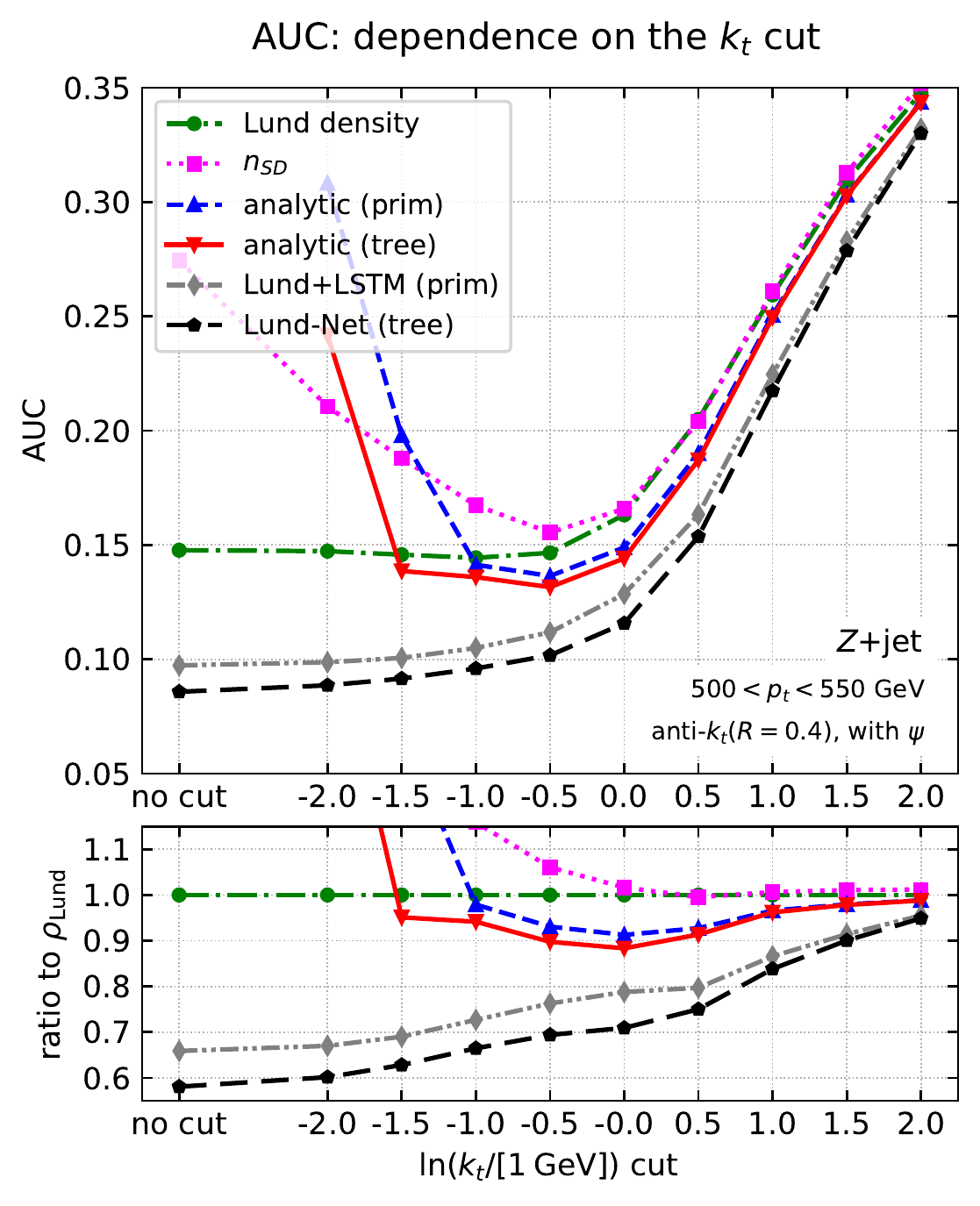}
  \caption{Area under the ROC curve (AUC) obtained using $Z$+jet
    events simulated with Pythia.
    All the Lund-plane based quark-gluon
    discriminants studied in the paper are shown: the Iterated Soft
    Drop multiplicity ($n_\text{SD}$), the likelihood based on the
    average primary Lund-plane density, and our new analytic and
    ML-based discriminants using either primary Lund branchings of the
    full Lund declustering tree.
    The AUC is plotted as a function of the minimum transverse
    momentum, $k_t$, cut imposed on the Lund declusterings.
  }\label{fig:full-auc}
\end{figure}

\begin{figure}
  \centering
  \begin{subfigure}[t]{0.48\textwidth}
    \includegraphics[width=\textwidth,page=1]{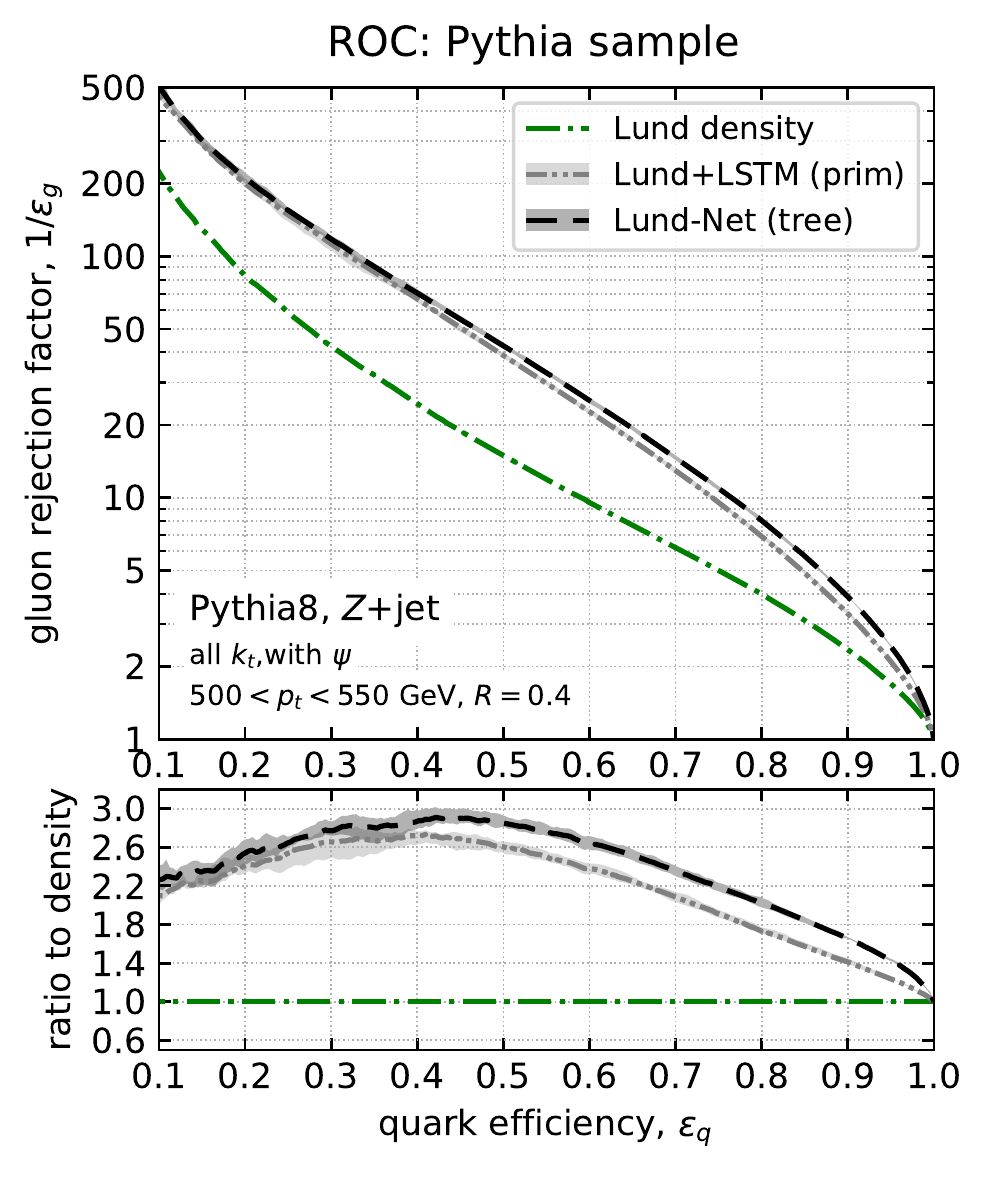}
    \caption{all $k_t$}\label{fig:full-roc-all}
  \end{subfigure}
  \hfill
  \begin{subfigure}[t]{0.48\textwidth}
    \includegraphics[width=\textwidth,page=5]{figs/plot-roc.pdf}
    \caption{$k_t>1$~GeV}\label{fig:full-roc-00}
  \end{subfigure}
  \caption{ROC curves corresponding to the AUC showed in
    Fig.~\ref{fig:full-auc}. The left plot corresponds to all Lund
    declustering being included while for the right plot a $k_t$ cut
    of 1~GeV has been applied. Since the methods anchored in
    perturbative QCD are meant to be effective in the region where
    perturbative QCD applies, they are only shown on the right
    plot.}\label{fig:full-roc}
\end{figure}

We first look at the performance of our taggers.
In this section we use our reference Monte Carlo sample, i.e.\ $Z$+jet
events generated with Pythia with hadronisation and multi-parton
interactions enabled.
Fig.~\ref{fig:full-auc} shows the area under the ROC curve as a
function of the $k_t$ cut applied on Lund declusterings, and
Fig.~\ref{fig:full-roc} shows the ROC curves themselves for two
specific choices of the cut: no cut (Fig.~\ref{fig:full-roc-all}) or
$k_t>1$~GeV (Fig.~\ref{fig:full-roc-00}).

Leaving aside for now the methods based on deep learning, we see the
expected pattern.
First, the average Lund density brings a small improvement compared to
the Iterated SoftDrop approach.
It is interesting to notice that while the performance of the Lund
density approach flattens as the $k_t$ cut is lowered, that of ISD
gets worse at small $k_t$ cuts.
Since the Iterated SoftDrop multiplicity and our analytic approaches
are based on perturbative QCD arguments, one might have anticipated
their performance degradation for low values of the $k_t$ cut. In
particular, since our analytic models include the running of the
strong coupling with $k_t$, they become unstable as we approach the
Landau pole. The average Lund density approach however directly uses
the Pythia sample to estimate the likelihood and is therefore free of
these effects.

Let us now focus on our analytic discriminants. Compared to the
average density we see a $\sim 5\%$ improvement in the AUC when using
only the primary declusterings, reaching $9\%$ for
$k_{t,\text{cut}}=1$~GeV. As visible in Fig.~\ref{fig:full-roc-00},
this improvement increases towards smaller quark efficiency where it
can reach $30{-}50\%$ for $\varepsilon_q$ in the $0.2{-}0.5$ range.
Adding the information from the full clustering tree, this improvement
in AUC increases slightly, reaches e.g.\ $\sim 12\%$ for
$k_{t,\text{cut}}=1$~GeV. Looking at the ROC curve, this improvement
is seen mostly at large $\varepsilon_q$ with limited impact at
smaller $\varepsilon_q$.
As for ISD, the performance of our analytic models worsens for small
$k_t$ cuts, below 1~GeV. This is most likely due to a breakdown of the
perturbative approach.

If we now turn to the Lund methods using deep learning, we see a clear
improvement in discriminating power for all $k_t$ cuts and across all
values of the quark efficiency. The AUC is reduced (i.e.\ improved) by
$20{-}40\%$ for a cut on $k_t$ below 1~GeV and the gluon rejection
factor is improved by a factor between 2 and 3 for $\varepsilon_q$ in
the $0.2{-}0.5$ range.

A striking feature of the machine-learning-based approaches in
Figs.~\ref{fig:full-auc} and~\ref{fig:full-roc}, especially compared
to the results shown in the collinear sample in
section~\ref{sec:mc-collinear}, is that they show a substantial
performance improvement compared to the analytic models.
There can be several explanations for this.
Of course, since our analytic approach is purely perturbative,
differences can be of non-perturbative origin. This is certainly the
case at very small values of $k_{t,\text{cut}}$ where out perturbative
approach breaks down when the performance of the
machine-learning-based approaches keeps improving.
However, the gain is already visible at values of the $k_t$ cut close
to 10~GeV, where non-perturbative corrections are relatively small.
From a pure perturbative perspective, there are at last three possible
explanations as to why the machine learning approaches may outperform
our analytic discriminant.

First, our treatment of clustering logarithms is only correct for
pairs of emissions at commensurate angles so we should expect
corrections even in the single-logarithmic limit. 

Secondly, our analytic discriminant works in the limit of small
angles, where quark-gluon discrimination can be thought of as
universal (at least within our single-logarithmic approximation). The
deep-learning methods will learn additional information, starting at
the single-logarithmic accuracy, from radiation at large angle. Since
this information is process-dependent, one should expect this gain
in performance to come at the expense of an enlarged sensitivity to
the hard process. We will come back to this point in
section~\ref{sec:mc-resilience}.

Lastly, there can be effects of subleading perturbative corrections
that are not included in our analytic approach. These can either be
subleading logarithmic corrections beyond single logarithms, or
finite, fixed-order, corrections which would induce additional
correlations between the Lund declusterings that are neglected at our
analytic accuracy but that the neural network training would pick.
In section~\ref{sec:asymptotic-tests}, we show that if we take a more
asymptotic regime, the gap between the analytic and deep-learning
approaches shrink, strongly suggesting that the differences seen in
Figs.~\ref{fig:full-auc} and~\ref{fig:full-roc} is not dominated by
our simplified treatment of clustering logarithms.

We should also point out that subleading logarithmic corrections,
or fixed-order corrections, are not fully included in Monte-Carlo
event generators like Pythia. The improvement seen with deep-learning
approaches should therefore be taken with caution.\footnote{It is
  tempting to argue that Monte Carlo event generators implement a more
  precise kinematics than the approximate one used in our analytic
  approach. For example, the Sudakov factor in the analytic
  calculations only retains the contributions up to single
  logarithms. For the fixed-coupling toy microjet sample used in
  section~\ref{sec:mc-collinear}, we had instead kept the full $z$
  dependence of the splitting function in the Sudakov factor. We have
  checked that the effect of keeping the full splitting instead of
  keeping only the terms relevant at the single-logarithmic accuracy
  is, at most, 0.5\%. This is clearly insufficient to explain the
  differences between the analytic and machine-learning approaches
  observed here.}

Of course, to this list of perturbative effects, one should also add
non-perturbative corrections which, while beyond the reach of our
analytic approach, are captured by the neural networks.

\subsection{Resilience}\label{sec:mc-resilience}

The discriminating power of a quark-gluon tagger is not necessarily
the only quality feature we may want to require.
Indeed, our extraction of the tagging performance is obtained for a
specific event sample which can have its own limitations or simply be
different from the event sample used in later practical applications.

Ideally, one would want a tagger to be {\em resilient}, i.e.\ to show
a degree of insensitivity to potential mismodelling aspects or to
specific details of an event sample.
For example, if we want to be able to describe a tagger from
first-principles perturbative QCD, we would want to limit its
sensitivity to the modelling of non-perturbative effects.
In a similar spirit, we want to limit the sensitivity of a tagger to
the details of the event generator used to obtain the event sample.
More specifically, in the context of quark-gluon discrimination, we
want our taggers to be insensitive to the details of the (hard)
processes contributing to the signal and background(s) we try to
separate.
This last point is intimately related to the intrinsic ill-defined
nature of quark-gluon tagging. In this context, resilience can be seen
as a measure of universality.\footnote{The idea of being resilient
  against details of the hard process however extends to tagging
  applications beyond quark-gluon discrimination.}

In this section, we therefore investigate the resilience of our tagger
against the three effects listed above: (i) non-perturbative effects,
(ii) the choice of the hard process and, (iii) the choice of a Monte
Carlo event generator.
The first is probed by comparing our reference sample to a sample
generated at parton level, i.e.\ with hadronisation and multi-parton
interactions switched off (see section~\ref{sec:mc-setup} for
details). For the second we use either our reference $Z+\text{jet}$
sample or a sample of dijet event, and for the third, we compare our
default Pythia8 sample to a Herwig7 sample.

For the analytic models, we apply them directly to the different event
samples, obtaining in each case the quark and gluon efficiency as a
function of the cut on the model's output, i.e.\ either the Iterated Soft
Drop multiplicity, or the analytic likelihood ratio for the ``Lund
density'' approach or for our new primary or full analytic
discriminants.
For machine-learning approaches, we have trained the networks on our
reference Pythia8 $Z+\text{jet}$ sample with hadronisation and
multi-parton interactions, and applied the resulting network to the
other event samples.

For a given fully-specified tagger, i.e.\ a tagging method and cut on
its output (a.k.a.\ a working point), one obtains quark and gluon
efficiencies $\varepsilon_{q,g}^\text{(ref)}$ and
$\varepsilon_{q,g}^\text{(alt)}$, respectively for the reference and
alternative event samples. The resilience of the taggers is then simply
defined as the inverse of the relative change between the two samples:
\begin{equation}\label{eq:resilience-definition}
  \zeta = \left[
    \left(
      \frac{2(\varepsilon_q^\text{(alt)}-\varepsilon_q^\text{(ref)})}
           {  \varepsilon_q^\text{(alt)}+\varepsilon_q^\text{(ref)} }
    \right)^2
  + \left(
      \frac{2(\varepsilon_g^\text{(alt)}-\varepsilon_g^\text{(ref)})}
           {  \varepsilon_g^\text{(alt)}+\varepsilon_g^\text{(ref)} }
     \right)^2
  \right]^{-1}.
\end{equation}
With this choice, a bigger $\zeta$ corresponds to a more resilient
tagger.
We should then select a working point at which
we evaluate the performance and resilience of a tagger.
In practice, we take the point at which the tagging performance,
defined as the significance $\varepsilon_q/\sqrt{\varepsilon_g}$, is
maximal for the reference sample. This is typically realised for
$\varepsilon\sim 0.3{-}0.5$.
The resulting performance (significance) and resilience are denoted by
$\Pi_\text{best}$ and $\zeta_\text{best}$, respectively.
Ideally, we therefore seek for a tagger with large $\Pi_\text{best}$ and
$\zeta_\text{best}$.
We have tested that selecting instead a fixed quark efficiency, e.g.\
$\varepsilon_q=0.5$, produces similar results.\footnote{In general,
  one can argue that the lower $\varepsilon_q$ values should be
  ignored because they are subject to large statistical
  fluctuations. The large and low $\varepsilon_q$ values are also
  impractical because they do not yield a large discriminating power.}
That said, while quantitative arguments can be made about the relative
discriminating performance of our taggers, it is more delicate to
reach such a precise quantitative discussion of resilience. The
discussion below therefore tries to remain mostly at a qualitative
level, i.e.\ noting that taggers with larger resilience are likely to
have less modelling uncertainties.
It would be interesting --- and clearly beyond the scope of this paper
--- to perform a dedicated study of resilience.

\begin{figure}
  \centering
  \begin{subfigure}[t]{0.48\textwidth}
    \includegraphics[width=\textwidth,page=1]{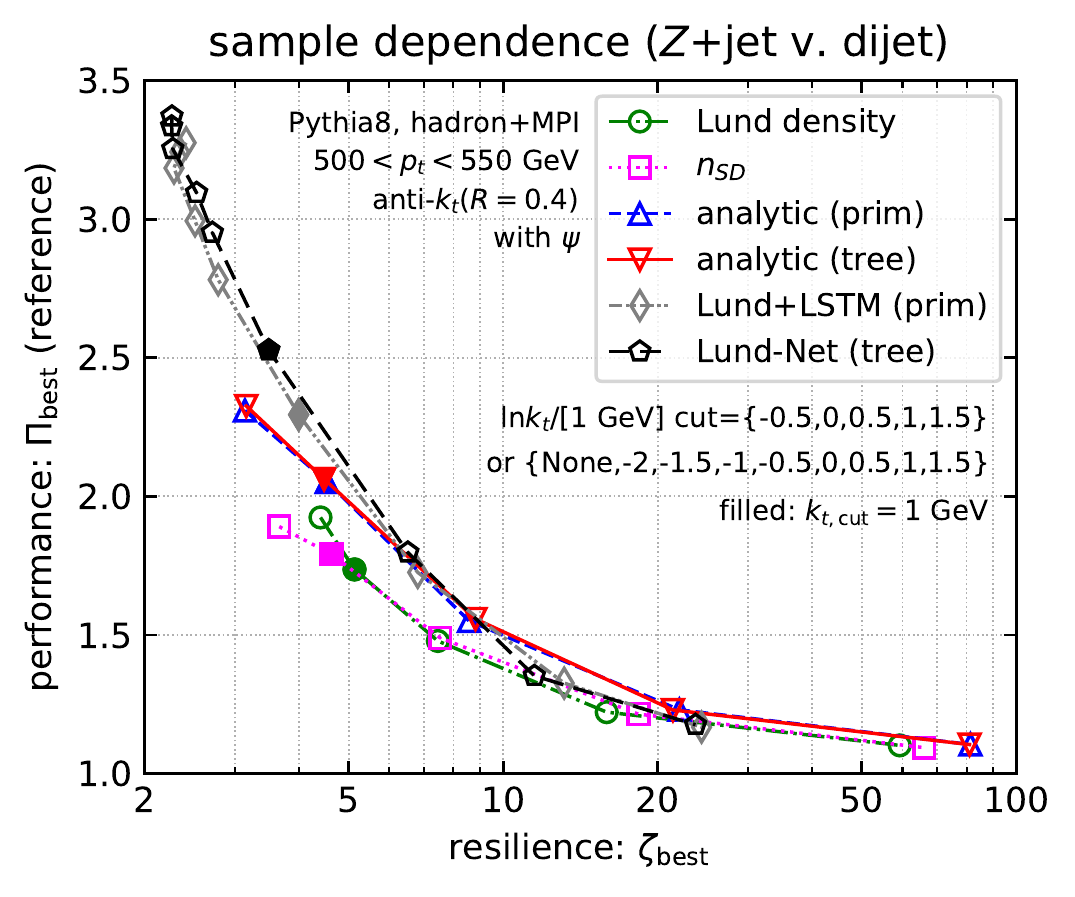}
    \caption{}\label{fig:resi-sample}
  \end{subfigure}
  \hfill
  \begin{subfigure}[t]{0.48\textwidth}
    \includegraphics[width=\textwidth,page=2]{figs/perf-v-resi-log.pdf}
    \caption{}\label{fig:resi-parton}
  \end{subfigure}\\
  \begin{subfigure}[t]{0.48\textwidth}
    \includegraphics[width=\textwidth,page=3]{figs/perf-v-resi-log.pdf}
    \caption{}\label{fig:resi-generator}
  \end{subfigure}
  %\hfill
  \caption{Plots of performance as a function of resilience for
    different discriminants. The curves are obtained by scanning over
    a range of accessible $k_{t,\text{cut}}$ values. The filled
    symbols correspond to a $k_t$ cut of 1~GeV. Different plots
    correspond to different probes of resilience: (a) probes the
    sample dependence (replacing the $Z$+jet sample with a dijet
    sample, (b) probes non-perturbative effects (using parton-level
    simulations instead of full simulations with hadronisation and
    multi-parton interactions) and (c) probes the effect of the Monte
    Carlo generator (using Herwig~7 instead of
    Pythia~8).}\label{fig:resi}
\end{figure}

We want to study how resilience and performance behave for our
discriminants, varying the $k_t$ cut on Lund declusterings.
Our results are presented in Fig.~\ref{fig:resi}, for the three types
of resilience we want to investigate: resilience against the specifics
of the hard process (Fig.~\ref{fig:resi-sample}), resilience against
non-perturbative effects (Fig.~\ref{fig:resi-parton}), and resilience
against the choice of the event generator
(Fig.~\ref{fig:resi-generator}).
To guide the eye, the results corresponding to a $k_t$ cut of 1~GeV
are represented with filled symbols, with all the other results using
open symbols.
For all three resiliences, the usual trade-off is observed: as we
increase the $k_t$ cut, performance decreases and resilience
increases.
Overall, our analytic models and our Deep Learning results show a
similar behaviour, although a given performance-resilience point is
achieved for a different $k_t$ cut for different taggers.
Our analytic models however appear as slightly more resilient to
non-perturbative effects than their machine-learning equivalents.

Compared to the Iterative Soft Drop and Lund density approaches,
one sees that the analytic model typically bring a gain in performance
without sacrificing in resilience.

Focusing on the results with a $k_t$ cut of 1~GeV, it is interesting
to see that the machine-learning-based techniques reach a larger
performance, as already seen in Figs.~\ref{fig:full-auc}
and~\ref{fig:full-roc}, at the expense of having a smaller
resilience.
This hints towards the interpretation that this gain in performance is
obtained by the neural networks exploiting information (i) going
beyond the ``universal'' collinear behaviour (worse resilience against
the choice of hard process), (ii) in non-perturbative effects (worse
resilience against hadronisation and MPI), and (iii) specific to the
modelling of the events (worse resilience against the choice of event
generator).
In all three cases, increasing the $k_t$ cut by a few hundred MeVs
would result in a behaviour very similar to the one of the analytic
model, both in terms of performance and in terms of resilience.

Finally, if all one cares about is performance, machine-learning
discriminants using the full information in the Lund tree show the
best result, albeit at the expense of a poor resilience.
This should at least be kept in mind when using a quark-gluon
discriminant for potentially different applications, or when assessing
uncertainties associated with a tagger.

\subsection{Comparison with other approaches}\label{sec:v-others}

In this section, we compare the performance of our Lund-plane-based
taggers to that of other existing taggers.

We therefore select our taggers based on the full Lund tree: the
Lund-Net tagger with no $k_t$ cut and the analytic discriminant
(``analytic(tree)'' in previous figures) with a $k_t$ cut of 1~GeV
(referred to as ``Lund NLL'' in this section), and compare them to a
series of pre-existing discriminants.
We first consider benchmark jet shapes:
\begin{itemize}
\item angularities~\cite{Berger:2003iw,Almeida:2008yp}, defined as the
  following sum over the jet constituents
  $\lambda_\alpha=(\sum_{i\in\text{jet}}p_{t,i}\Delta
  R_{i,\text{jet}}^\alpha)/(R^\alpha\sum_{i\in\text{jet}}p_{t,i})$. We
  work with $\alpha=1$, sometimes referred to as width or girth.
\item energy-energy correlation functions~\cite{Larkoski:2013eya},
  defined as the following sum over pairs of jet constituents
  $\text{EEC}_\beta=(\sum_{i,j\in\text{jet}}p_{t,i}p_{t,j}\Delta
  R_{ij}^\beta)/[R^\beta(\sum_{i\in\text{jet}}p_{t,i})^2]$.
  In this case, we will set $\beta=1/2$. 
\item to probe the effect of a $k_t$ cut similar to the one we
  introduce in the Lund plane techniques, we have considered the case
  where the angularities and energy-correlation function are defined
  on the Lund declusterings (primary and secondary) above a given
  $k_t$ cut.
  We recall that in this case, we expect that the Iterative Soft Drop
  multiplicity and our analytic Lund-tree discriminant are respectively
  optimal at leading and next-to-leading logarithmic accuracy in QCD.
\end{itemize}
We then consider a series of recent machine-learning-based quark-gluon
discriminants:
\begin{itemize}
\item Particle-Net described in Ref~\cite{Qu:2019gqs}, based on point
  clouds.
  In practice, we have directly used the {\tt ParticleNet} code available
  from~\cite{particle-net-code}, modifying the provided {\tt keras}
  example to use our event sample. We have used a batch size of 1000
  and kept the best model over a training of 50 epochs. Note that this
  model includes the particle ID in the network inputs.
\item Particle-Flow Networks (PFN) from Ref.~\cite{Komiske:2018cqr}, 
  This includes the rapidity, azimuth (both relative to the jet axis)
  and transverse momentum information of each jet constituent. Each
  particle is mapped into a per-particle latent space. The sum over
  all particles of these spaces is then mapped onto a final
  discriminating variable.
  We have also considered the so-called PFN-ID approach where the
  particle IDs are also included.
  In practice, we have used the code provided in the {\tt EnergyFlow}
  package~\cite{energy-flow-code}, modifying the examples to use our
  event samples and training over 60 epochs.
\item Energy-Flow Networks (EFN) also from~\cite{Komiske:2018cqr} and
  again only adapted from the example given in the {\tt EnergyFlow}
  package to our needs.
  The approach is similar to that of the PFN above except that the
  latent space uses IRC-safe information through a weighting
  proportional to the $p_t$ of each particle.
\end{itemize}
Finally, we also consider the Lund-Net approach, labelled
``Lund-Net(+ID)'' where each Lund tuple $(\ln \Delta, \ln k_t,
\ln z, \Psi)$ is supplemented by one
additional integers for each of the two subjets $j_1$ and $j_2$ and
determined as follows: if the subjet has a single constituent we use
the PDG ID of the constituent, otherwise we set the integer to 0.
The idea is similar to having particle identification added from the
PFN to the PFN-ID approach.

\begin{figure}
  \centering
  \begin{subfigure}[t]{0.485\textwidth}
    \includegraphics[width=\textwidth, page=2]{figs/plot-v-others.pdf}
    \caption{}\label{fig:v-others-sig-analytic}
  \end{subfigure}
  \hfill
  \begin{subfigure}[t]{0.485\textwidth}
    \includegraphics[width=\textwidth, page=1]{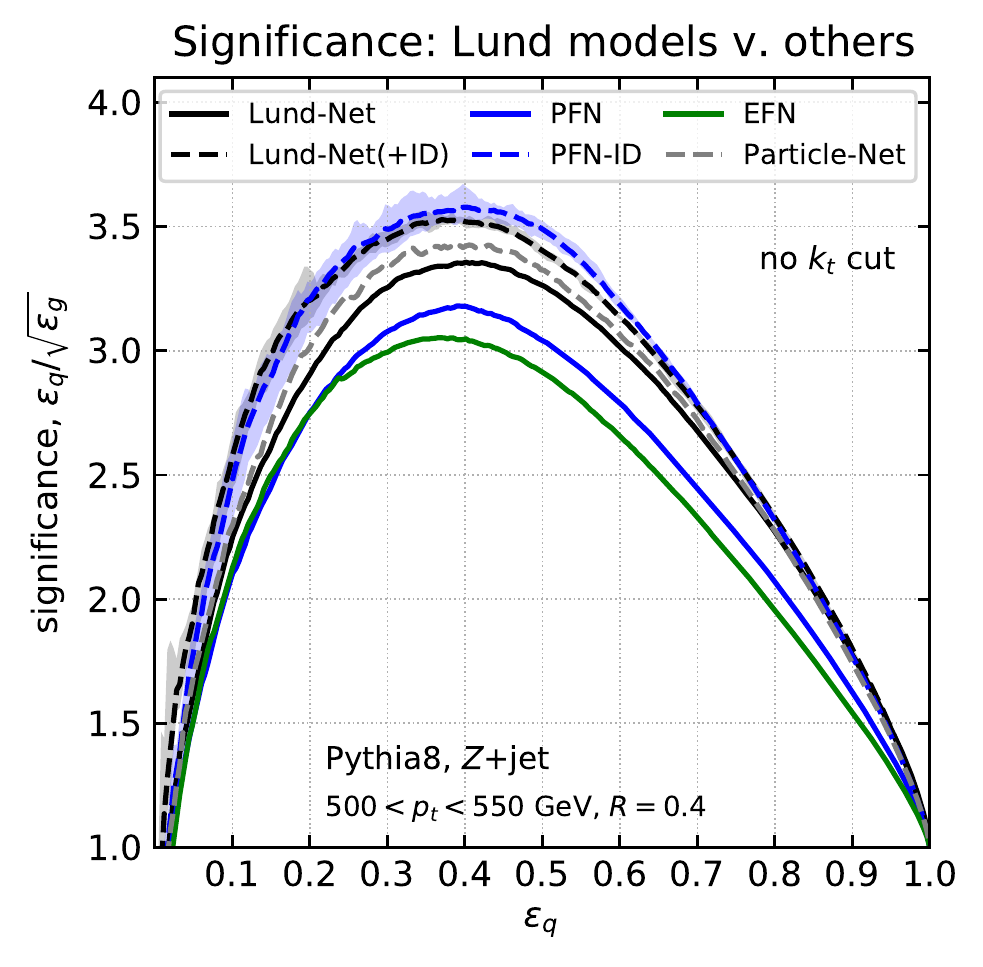}
    \caption{}\label{fig:v-others-sig-ml}
  \end{subfigure}\\
  \begin{subfigure}[t]{0.475\textwidth}
    \adjustbox{valign=t}{
      \includegraphics[width=\textwidth, page=3]{figs/plot-v-others.pdf}
    }
    \caption{}\label{fig:v-others-resi}
  \end{subfigure}
  \hfill
  \begin{subfigure}[t]{0.485\textwidth}
    \adjustbox{valign=t}{
      \begin{tabular}{l|l}
        model & AUC \\
        \hline
        $n_{\rm SD}$ {\scriptsize ($k_t\!>\!1$ GeV)} & 0.1658 \\
        Lund NLL {\scriptsize ($k_t\!>\!1$ GeV)}   & {\bf 0.1441} \\
        EEC$_{0.5}$ {\scriptsize (all $k_t$)}   & 0.2074 \\
        EEC$_{0.5}$ {\scriptsize ($k_t\!>\!1$ GeV)} & 0.2150 \\
        $\lambda_1$ {\scriptsize (all $k_t$)}   & 0.2270 \\
        $\lambda_1$ {\scriptsize ($k_t\!>\!1$ GeV)} & 0.2371 \\
        \hline
        Lund-Net                & 0.0858 $\pm$ 0.0007 \\
        Lund-Net(+ID)           & {\bf 0.0835 $\pm$ 0.0005} \\
        Particle-Net            & 0.0871 $\pm$ 0.0009 \\
        PFN                     & 0.0994 $\pm$ 0.0009 \\
        PFN-ID                  & 0.0853 $\pm$ 0.0005 \\
        EFN                     & 0.1080 $\pm$ 0.0010 \\
      \end{tabular}
   }
  \end{subfigure}
  \caption{Comparison of the Lund-plane-based approaches with other
    models. Explicit plots of the signal significance
    $\varepsilon_q/\sqrt{\varepsilon_g}$ are shown in the upper plots,
    first for analytic discriminants, figure (a), then for
    machine-learning-based approaches, figure (b). The bottom panel,
    figure (c), shows the corresponding performance v.\ resilience
    plot, where the resilience is measured with respect to the choice
    of Monte Carlo generator
    (cf.~section~\ref{sec:mc-resilience}).
    The table on the bottom-right corner gives the area under the ROC
    curve (AUC) for the different models (lower is batter). For the
    ML-based models, the uncertainty is half the difference between
    the minimal and maximal values obtained over 5 different runs.
  }\label{fig:v-others}
\end{figure}

Our findings are presented in Fig.~\ref{fig:v-others}, for the signal
significance (top row), and for the trade-off
between significance and resilience against the choice of
Monte-Carlo event generator (bottom row).
Focusing first on the signal significance for analytic discriminants,
Fig.~\ref{fig:v-others-sig-analytic}, we see relatively different
patterns between the shape-based observables $\lambda_1$ and
$\text{EEC}_{0.5}$ and the Lund-based observables, with the
performance peaking at larger values of the quark efficiency in the
latter case, with shape-based observables reaching a better overall
performance.
However, computing the EEC and $\lambda$ using the Lund declusterings
with a $k_t$ above 1~GeV, i.e.\ with the same input information in all
cases, we see that our analytic Lund discriminant shows indeed an
improvement at all quark efficiencies compared to the shape-based
discriminants.
A similar pattern is seen in the resilience plot on
Fig.~\ref{fig:v-others-resi} with the Lund analytic model with a 1-GeV
$k_t$ cut being intermediate between the jet shape with a 1-GeV $k_t$
cut and with no $k_t$ cut.
It is interesting to notice that the addition of particle ID
information to the Lund-Net approach improves the performance at low
$k_t$ cut, or with no $k_t$ cut at all, but changes neither the
performance nor the resilience once a larger $k_t$ cut is
applied. This is most likely due to the fact that all the input
subjets have more than one constituent and hence the ID information is
0. This contrasts with the findings in Ref.~\cite{Dreyer:2020brq},
where the addition of the jet mass had a negative impact on resilience
at large $k_t$ cuts. 
At low resilience (large significance), the jet shapes give a slightly
better performance vs.\ significance behaviour than our Lund-plane
approach.
Focusing instead on the AUC --- the bottom-right table in
Fig.~\ref{fig:v-others} --- we see that our analytic Lund-tree
approach does a better job than the other jet shapes (i.e. a lower
AUC), including jet shapes computed on the full set of constituents.

Moving to machine-learning-based models,
Fig.~\ref{fig:v-others-sig-ml}, we see a significance pattern mostly
similar across different models. The performance of Lund-based models
is on par with the one obtained from Particle-Net. Compared to
energy/particle flow approaches, our Lund-based results show a
slightly better performance than the EFN and PFN results, but fall
slightly lower than the performance of PFN-ID. Adding the particle ID
information using our Lund-Net(+ID) approach recovers a performance
similar to the the PFN-ID approach, although with a marginally smaller
average peak performance.
If we instead look at the AUC, we see that the Lund-Net(+ID) reaches
the best performance (lowest AUC), marginally better that the PFN-ID
and Lund-Net approaches, then followed by the Particle-Net model.
While the PFN-ID method shows a small performance improvement at mid
signal efficiency, the Lund-Net(+ID) setup has a small advantage at
small and large signal efficiencies.\footnote{Including the particle
  ID information in a more coherent way, e.g.\ as a separate
  information that is fed to the final dense layers, one might be able
  to make up the difference with PFN-ID at mid quark efficiencies.}
The observed differences are however of a size similar to the
statistical fluctuations observed in our simulations (only shown, on
the significance plot, for the Lund-Net(+ID) and PFN-ID for the sake
of readability).

Finally, from Fig.~\ref{fig:v-others-resi}, we see a similar degree of
resilience for all machine-learning-based approaches.
Again, it would be interesting to train energy/particle flow networks
or the ParticleNet network on Lund declusterings above a certain $k_t$
cut (or using another cut definition) to study the
performance versus resilience trade-off in a broader perspective.

\subsection{Effect of clustering logarithms and of azimuthal
  angles}\label{sec:clust-azimuth-effects}

All the results presented so far have included the dependence on the
Lund azimuthal angles $\psi_i$.
% T
Since it is known that these are not properly described at
single-logarithmic accuracy by the standard dipole showers (including
Pythia8)~\cite{Dasgupta:2020fwr}, we want to briefly investigate their
impact on discriminating power.
Additionally, from our analytic perspective, the dependence on the
Lund azimuthal angles $\psi_i$ only comes in through the clustering
logarithms.
This therefore gives us an explicit opportunity to investigate the
numerical impact of including clustering logarithms into our analytic
discriminants.

\begin{figure}
  \centering
  \includegraphics[width=0.9\textwidth,page=1]{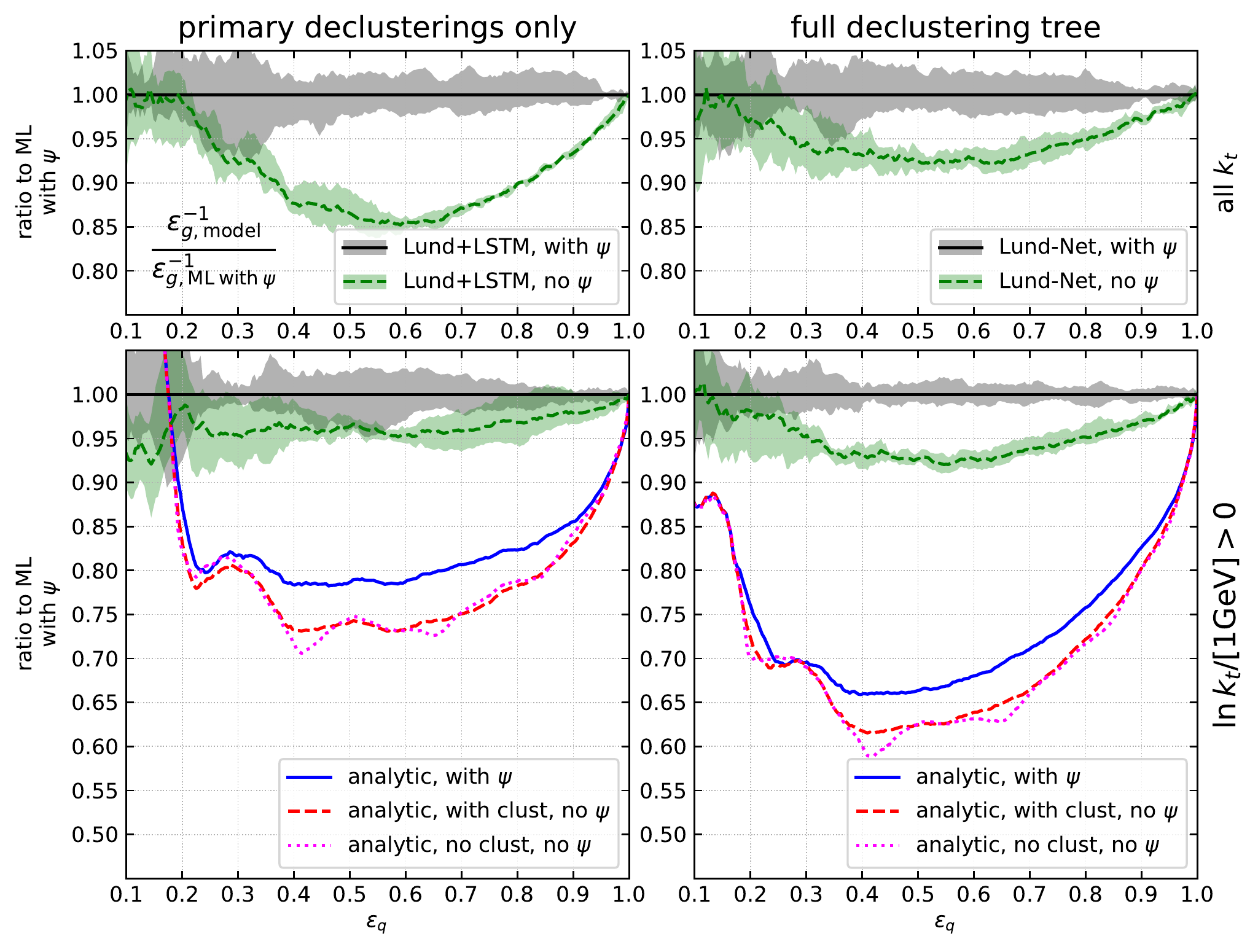}
  \caption{Plots showing the ratio of the ROC curve with and without
    azimuthal-angle dependence and/or clustering logarithms. The plots
    of the left column include only primary declusterings while the
    plots of the right column include the full declustering tree. The
    bottom plots include declusterings above a 1-GeV $k_t$ cut, while
    the top plots include all relative transverse momenta. In all
    cases, the ROC curve is normalised to the ML Results including
    azimuthal-angle dependence.}\label{fig:clusteting-psi}
\end{figure}

In Fig.~\ref{fig:clusteting-psi}, we show the background rejection,
$1/\varepsilon_g$, including either only primary declusterings (left
column) of the full declustering tree (right column), as obtained
using our reference Pythia~8 $Z$+jet sample with hadronisation and
multi-parton interactions.
The top row, with only machine-learning results, corresponds to the
case without a $k_t$ cut while the bottom row, with both ML and
analytic results, includes only the declusterings with $k_t\ge 1$~GeV.
To increase readability, we show in all cases, the ratio relative to
what is obtained with the corresponding ML model --- LSTM or Lund
depending on whether only the primary or all the declusterings are
used --- including the $\psi$ dependence.

We first discuss the ML results, presented in
Fig.~\ref{fig:clusteting-psi} either with (solid, black) or without
(dashed, green) $\psi$ information. We see that including the $\psi$
information brings a $5{-}15$\% performance gain, mostly at
intermediate quark efficiency. This gain is larger at lower $k_t$
where non-perturbative effects are larger.
When a 1-GeV $k_t$ cut is imposed, we also show the results of our
analytic quark-gluon taggers, again using either primary-only
information (left column), or using the full declustering tree (right
column). In each case, three results are given: including the $\psi$
angles (solid, blue), not including the $\psi$ angles but including
the ($\psi$-averaged) clustering logarithms from
Eq.~(\ref{eq:omega-average}) (dashed, red), or including neither the
$\psi$ angles, nor the clustering logarithms (dotted, magenta).
We again see a $\sim 10$\% increase in performance brought by the
inclusion of azimuthal angles.
The fact that these performance gains are of similar magnitude in the
deep-learning and analytic approaches indicates that our simplified
treatment of clustering logarithms is a decent approximation.

Finally, when the azimuthal angles are not included, we see that the
influence of clustering logarithms is small.

\subsection{Asymptotic single-logarithmic limit}\label{sec:asymptotic-tests}

In this final study, we want to further study the differences between
the analytic and ML results.
Our aim is here to take a limit where subleading effects decrease.
Since our analytic approach technically resums double and single
logarithms of $\log(p_tR/k_{t,\text{cut}})$, we want to proceed in a
similar way as for the NLL-accuracy tests in~\cite{Dasgupta:2020fwr},
i.e,\ take the limit $\alpha_s(p_tR)\to 0$, $\log(p_tR/k_{t,\text{cut}})\to
\infty$ while keeping $\alpha_s(p_tR)\log(p_tR/k_{t,\text{cut}})$
constant.
The main idea behind this limit is that subleading-logarithmic
contributions as well as fixed-order contributions are suppressed as
$\alpha_s\to 0$.

Generating and analysing event over an exponentially increasing range
of scales poses a series of numerical challenges which, in practice,
make it unreachable for standard Monte Carlo event generators like
Pythia8.
We have therefore used instead the PanScales $e^+e^-$ code developed
precisely to overcome these challenges in
Ref.~\cite{Dasgupta:2020fwr}.
We have therefore generated $e^+e^-\to Z\to q\bar q$ (quark) events
and $e^+e^-\to H\to gg$ (gluon) events with a centre-of-mass energy
$Q$, fixing $\alpha_s(Q)\log(Q/k_{t,\text{cut}})=0.32$ and taking
$\alpha_s(Q)$ to be either 0.04, 0.02, 0.01, corresponding to
$L=\log(Q/k_{t,\text{cut}})$ of either 8, 16, or 32.
In all cases, we have used the PanLocal shower in its antenna variant
with the $\beta$ parameter set to $1/2$. Subleading colour corrections
are included using the Nested Ordered Double-Soft (NODS) scheme as
described in~\cite{Hamilton:2020rcu}. This produces event samples with
a single-logarithmic accuracy with the exception of the
subleading-$N_c$ corrections for which the NODS method only guarantees
the correct behaviour for (any number of) pairs of emissions at
commensurate angles.
We reconstruct the $e^+e^-$ Lund declusterings in each of the event
hemispheres. The $e^+e^-$ reconstruction follows an almost-trivial
adaptation of the hadronic collisions technique described in
section~\ref{sec:earlier}, except perhaps for the reconstruction of
the azimuthal angle $\psi$ which is described in details
in~\cite{Dasgupta:2020fwr}.

\begin{figure}
  \begin{subfigure}[t]{0.46\textwidth}
    \includegraphics[width=\textwidth,page=1]{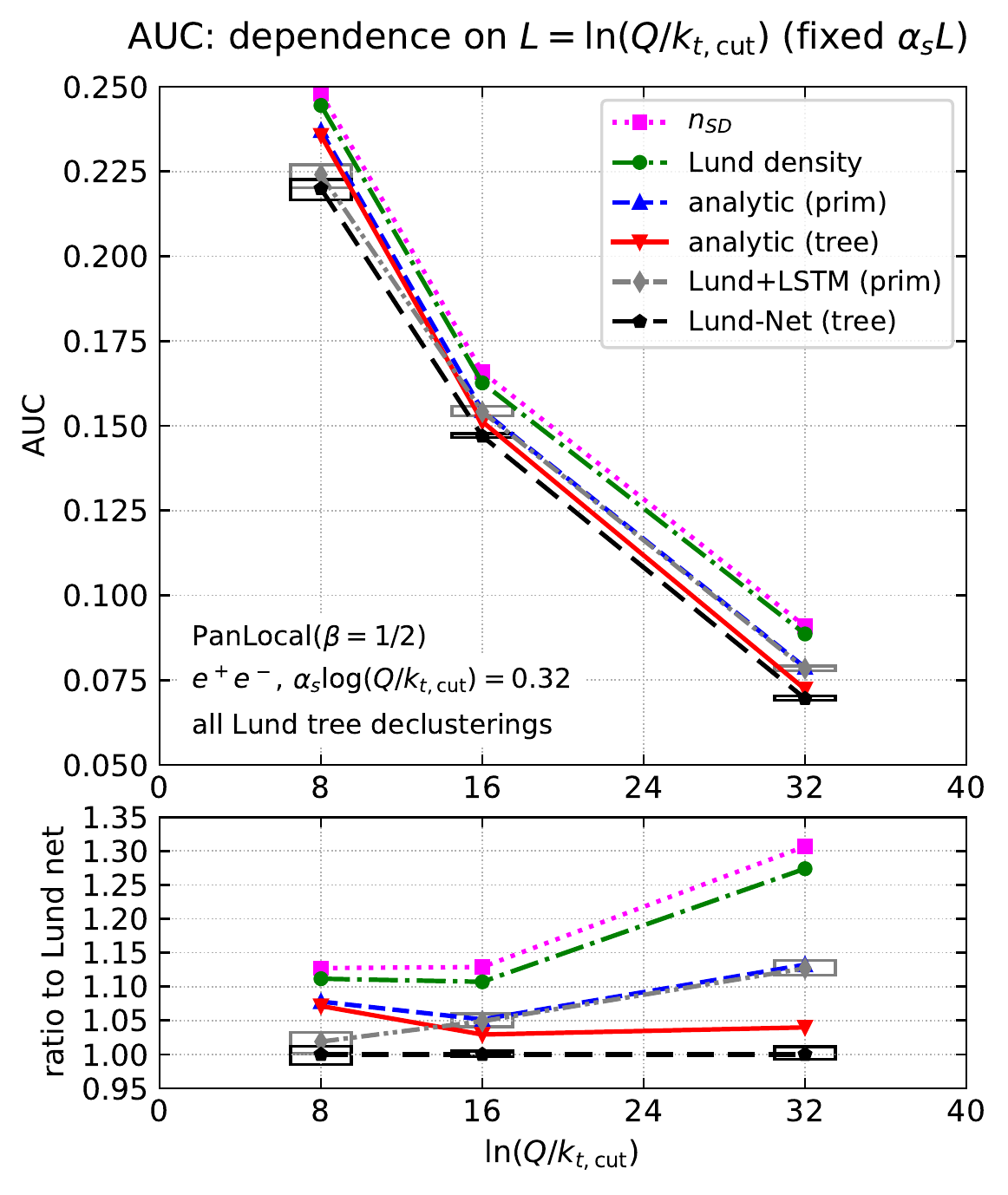}
    \caption{}\label{fig:asympt-auc}
  \end{subfigure}
  \hfill
  \begin{subfigure}[t]{0.5\textwidth}
    \includegraphics[width=\textwidth,page=2]{figs/plot-panscales-asympt.pdf}
    \caption{}\label{fig:asympt-roc}
  \end{subfigure}\\
  \caption{Dependence of (a) the AUC, and (b) the ROC curve on
    $L=\log(Q/k_{t,\text{cut}})$ fixing $\alpha_sL=0.32$. The results
    are obtained with an $e^+e^-$ parton-level setup using the
    PanLocal PanScales shower which is NLL-accurate. Different curves
    show different methods and we see that the analytic and ML models
    converge to one another as we tale the asymptotic limit $L\to
    \infty$, $\alpha_s\to 0$, $\alpha_sL=\text{constant}$. For the
    plots on the right column, we take the ratio of the ROC curves to
    the Lund-Net results.}\label{fig:asympt}
\end{figure}

Both our analytic methods and the approaches based on Machine Learning
can be straightforwardly applied to these new sets of events. We
therefore study the same set of quark-gluon discriminants as in
section~\ref{sec:mc-performance}.
The contribution from clustering logarithms is included in our
analytic models and information from the Lund azimuthal angles is
included both in the analytic and ML approaches.

Our results are presented in Fig.~\ref{fig:asympt} for the AUC,
Fig.~\ref{fig:asympt-auc} and for the ROC curves,
Fig.~\ref{fig:asympt-roc}. For the latter, we have normalised the
gluon rejection rate $\varepsilon_g^{-1}$ to the Lund-Net rejection
rate.
One can see from these plots that the difference between the analytic
and Machine-Learning-based methods decrease when increasing $L$
(decreasing $\alpha_s$) at fixed $\alpha_sL$. This is true separately
for the approaches using only primary declusterings
(``analytic(prim)'' and ``Lund+LSTM'') and for the approaches using
the full declustering tree (``analytic(tree)'' and ``Lund-Net'').
At the same time, the performance gain compared to the Iterated
SoftDrop multiplicity increases.

Before closing this section, we want to address a last point about the
azimuthal angle dependence and single-logarithmic accuracy.
It has been pointed out in Ref.~\cite{Dasgupta:2020fwr} that, due to
non-physical recoil effects, dipole showers such as Pythia or Dire, would
generate a spurious dependence on $\psi$, potentially biasing the
assessment of quark-gluon classification.
Using the machinery described in this section, we have studied
potential differences between the Pythia shower and the PanScales
showers which are free of this effect~\cite{Dasgupta:2020fwr}.
Within the few-percent accuracy of our studies we have not been able
to isolate a clearly-visible impact of this effect.

%======================================================================
\section{Conclusions}\label{sec:concl}

This paper addresses the question of quark/gluon discrimination using
the Lund-plane approach to characterise the substructure of jets.
Our main result is that it is possible to compute the quark-gluon
likelihood ratio from the first principles in QCD. The calculation is done
at the single-logarithmic accuracy, including all collinear
contributions as well as clustering effects for any number of pairs of
emissions at commensurate angles.
This automatically provides us with an optimal quark/gluon tagger at
the same accuracy.

As expectable, this tagger shows an improved performance either
compared to using the average Lund plane density to build the
likelihood ratio, or compared to the Iterated Soft Drop multiplicity
which corresponds to the optimal quark/gluon discriminant at leading
(double) logarithmic accuracy.
Most of the improvement ($\lesssim 10$\% for the AUC compared to
$n_\text{SD}$) is already captured when including only primary
declusterings, but the effect of additional declustering in
subsidiary Lund planes ($\sim 3$\%) is clearly visible, especially at
larger quark efficiencies.
The gain in performance can be attributed to the better treatment of
the kinematics of each emission, e.g.\ through the full
Altarelli-Parisi splitting functions, through the full antenna
pattern for emissions at commensurate angles, and to a better
treatment of the correlations between emissions, e.g.\ by taking into
account the energy of the emitting parton or by including clustering
effects.
Furthermore, this gain in performance is accompanied by a gain in
resilience against effects beyond our perturbative calculation. In
this context, we have studied three specific effects: the dependence
against non-perturbative effects, the dependence against the specific
choice of quark/gluon enriched samples used as benchmarks, and the
choice of Monte Carlo event generator.

In Section~\ref{sec:v-others}, we have compared our Lund-based
approach to other typical quark/gluon taggers using jet substructure,
like angularities or energy correlation functions. Focusing for
simplicity on the case where all the taggers are applied to Lund
declusterings with a $k_t$ above 1~GeV to reduce non-perturbative
effects, we see that the Lund-based likelihood approach gives a gain
in performance, especially at large quark efficiency, while
maintaining a similar degree of resilience.\footnote{If the jet shapes
  are computed on the full jet, they yield a larger significance at
  smaller quark efficiency at the expense of a reduced resilience
  against non-perturbative effects.}

The second set of results in this paper is the extensive study of
quark/gluon taggers using deep-learning techniques combined with
Lund declustering information.
When applied to the full set of declusterings in a jet, our ML-based
tagger reaches a performance (and resilience) comparable to that
obtained with Particle-Flow networks~\cite{Komiske:2018cqr}, and marginally better than what
is achieved by Particle-Net~\cite{Qu:2019gqs}.
Compared to the analytic results, our ML-tagger gives a clearly visible
performance gain, even when considering declusterings above a given
$k_t$ cut.
This gain in performance, however, comes at a price in all the forms of
resilience we have studied, especially the sensitivity to
hadronisation and multi-parton interactions.

One of the key points of this paper is the direct comparison between
the analytic and deep-learning approaches. Since our tagger targets
the optimal discriminant in the single-logarithmic approximation one
can directly compare its performance with that of the deep-learning
models.
We first did that in the strongly angular-ordered limit where our
analytic calculation is exact.
The results in section~\ref{sec:mc-collinear} indicate a convergence
of the deep-learning taggers to the optimal performance as long as the
size of the network is taken large enough.
Beyond the collinear limit, where our analytic treatment is only
approximate, the deep-learning approach shows a better
performance. However, if we move progressively to the
single-logarithmic asymptotic limit ($\alpha_s\to 0$ at fixed
$\alpha_s\log(Q/k_{t,\text{cut}}$) the difference between the two
approaches drastically reduces as we showed in section~\ref{sec:asymptotic-tests}. At the same time, the gain in
performance compared to leading-logarithmic-accurate taggers --- i.e.\
the average Lund density and the Iterated Soft Drop multiplicity ---
increases.

The above observations strongly suggest that the gain in performance
observed for deep-learning taggers (in addition to our analytic
tagger) in phenomenological Monte Carlo applications come
predominantly either from subleading effects (beyond single
logarithms), from large-angle soft emissions (not included in our
analytic calculation), or from non-perturbative effects.

This points towards several interesting physics considerations.
First, subleading logarithmic effects, albeit present in data, are not
properly included in any parton shower Monte Carlo generator today.
Conclusions regarding subdominant logarithmic effects should
therefore be taken carefully.
In this context, it would be interesting in the future to further
investigate potential differences between standard dipole showers
(like Pythia) which are known to have failures at the single-logarithmic
accuracy, or even at leading-log for subleading colour effects, and
the PanScales showers which are NLL-accurate.

Then, large-angle soft emissions are process-dependent and should
therefore be treated carefully when applied outside the configurations
where they have been tested and calibrated.
In the future, it would be interesting to see if an analytic treatment
similar to the one adopted in this paper could allow for quantitative
assessment of the process-dependence of quark/gluon discrimination
(see also Ref.~\cite{Bright-Thonney:2018mxq} for a Monte-Carlo-based
study).

Finally, non-perturbative effects come with non-negligible modelling
uncertainties and should therefore also be taken carefully.
The ability to progressively reduce non-perturbative effects by
increasing the $k_t$ cut-off on Lund declusterings could help further
investigating the impact of non-perturbative effects, and the
associated systematic uncertainties, in a practical context.

In conclusion, we have seen that Lund-plane declusterings were useful
to define a variety of quark/gluon discriminants, bridging regions
targetting high discriminating performance and regions where
a high-precision degree of control can be reached from
first-principles QCD.

%======================================================================
\section*{Acknowledgements}
G.S.\ is especially grateful to Ben Nachman and Eric Metodiev for
stimulating discussions.
G.S.\ and F.D.\ are also grateful to Gavin Salam for earlier work on
the Lund jet plane without which this paper would never have been possible.
A.T.\ wishes to thank the Institut de Physique Theorique for the
hospitality.
This work has been supported in part by the French Agence Nationale de
la Recherche, under grant ANR-15-CE31-0016 (G.S. and A.T.), by the
European Research Council (ERC) under the European Union’s Horizon
2020 research and innovation programme (grant agreement No.\ 788223,
PanScales) (G.S.), by the Starting Grant from Trond Mohn Foundation
(BFS2018REK01) (A.T.), and by a Royal Society University Research
Fellowship (URF$\backslash$R1$\backslash$211294) (F.D.).

%======================================================================
\appendix

%======================================================================
\section{Sudakov factors with exact splitting functions}\label{sec:sudakov-full-splitting}

In section~\ref{sec:mc-collinear}, we have used an event sample
generated in the strong-angular-ordered limit to compare the
performance of our analytic discriminants (also in the limit of strong
angular ordering, see section~\ref{sec:discrim-analytic} to that of
deep-learning-based discriminants.
To guarantee that the analytic approach reproduces the exact
likelihood ratio, we have kept the full Altarelli-Parisi splitting
functions in the Sudakov (using a fixed-coupling approximation).
If we have a hard parton of momentum $x p_t$ (with $p_t$ the initial
transverse momentum of the jet) and flavour $f$, and compute the
Sudakov factor between an angle $\Delta_{i-1}$ and $\Delta_i$, we find:
\begin{equation}
  -\log S_f^{(i-1,i)} = 
\frac{2 \alpha_s C_f}{pi}  \left[\frac{\log^2 x_1}{2} - \frac{\log^2
  x_2}{2} + B_f \log\frac{x_2}{x_1} +
  \text{Li}_2(x_1)-\text{Li}_2(x_2) + \delta R_f\right],
\end{equation}
with
\begin{align}
  x_1 & = \frac{k_{t,\text{cut}}}{x\Delta_{i-1}p_t},
  & B_q &=-\frac{3}{4}, \\
  x_2 & = \frac{k_{t,\text{cut}}}{x\Delta_ip_t},
  & B_g &=-\frac{11C_A-4n_f T_R}{12 C_A},
\end{align}
and
\begin{align}
  \delta R_q &= \frac{3}{2}(x_2-x_1), \\
  \delta R_g &= \frac{3}{2}(x_2-x_1) + \left(\frac{1}{2}-\frac{n_fT_R}{C_A}\right)\left[(x_2-x_1)-\frac{1}{2}(x_2^2-x_1^2)+\frac{2}{9}(x_2^3-x_1^3)\right].
\end{align}

%======================================================================
\bibliographystyle{JHEP}
\bibliography{refs}

\providecommand{\href}[2]{#2}\begingroup\raggedright\begin{thebibliography}{10}

\bibitem{Salam:2009jx}
G.~P. Salam, \emph{{Towards jetography}},
  \href{https://doi.org/10.1140/epjc/s10052-010-1314-6}{\emph{Eur. Phys. J.}
  {\bfseries C67} (2010) 637}
  [\href{https://arxiv.org/abs/0906.1833}{{\ttfamily 0906.1833}}].

\bibitem{Marzani:2019hun}
S.~Marzani, G.~Soyez and M.~Spannowsky, \emph{{Looking inside jets: an
  introduction to jet substructure and boosted-object phenomenology}},
  vol.~958. Springer, 2019,
  \href{https://doi.org/10.1007/978-3-030-15709-8}{10.1007/978-3-030-15709-8},
  [\href{https://arxiv.org/abs/1901.10342}{{\ttfamily 1901.10342}}].

\bibitem{Gras:2017jty}
P.~Gras, S.~Hoeche, D.~Kar, A.~Larkoski, L.~L{\"o}nnblad, S.~Platzer et~al.,
  \emph{{Systematics of quark/gluon tagging}},
  \href{https://doi.org/10.1007/JHEP07(2017)091}{\emph{JHEP} {\bfseries 07}
  (2017) 091} [\href{https://arxiv.org/abs/1704.03878}{{\ttfamily
  1704.03878}}].

\bibitem{Komiske:2018vkc}
P.~T. Komiske, E.~M. Metodiev and J.~Thaler, \emph{{An operational definition
  of quark and gluon jets}},
  \href{https://doi.org/10.1007/JHEP11(2018)059}{\emph{JHEP} {\bfseries 11}
  (2018) 059} [\href{https://arxiv.org/abs/1809.01140}{{\ttfamily
  1809.01140}}].

\bibitem{Larkoski:2019nwj}
A.~J. Larkoski and E.~M. Metodiev, \emph{{A Theory of Quark vs. Gluon
  Discrimination}}, \href{https://doi.org/10.1007/JHEP10(2019)014}{\emph{JHEP}
  {\bfseries 10} (2019) 014}
  [\href{https://arxiv.org/abs/1906.01639}{{\ttfamily 1906.01639}}].

\bibitem{Banfi:2006hf}
A.~Banfi, G.~P. Salam and G.~Zanderighi, \emph{{Infrared-safe definition of jet
  flavour}}, {\emph{Eur. Phys. J.} {\bfseries C47} (2006) 113}
  [\href{https://arxiv.org/abs/hep-ph/0601139}{{\ttfamily hep-ph/0601139}}].

\bibitem{Larkoski:2017jix}
A.~J. Larkoski, I.~Moult and B.~Nachman, \emph{{Jet Substructure at the Large
  Hadron Collider: A Review of Recent Advances in Theory and Machine
  Learning}},  \href{https://arxiv.org/abs/1709.04464}{{\ttfamily 1709.04464}}.

\bibitem{Asquith:2018igt}
L.~Asquith et~al., \emph{{Jet Substructure at the Large Hadron Collider :
  Experimental Review}},  \href{https://arxiv.org/abs/1803.06991}{{\ttfamily
  1803.06991}}.

\bibitem{jet_images2}
L.~de~Oliveira, M.~Kagan, L.~Mackey, B.~Nachman and A.~Schwartzman,
  \emph{{Jet-images -- deep learning edition}},
  \href{https://doi.org/10.1007/JHEP07(2016)069}{\emph{JHEP} {\bfseries 1607}
  (2016) 069} [\href{https://arxiv.org/abs/1511.05190}{{\ttfamily
  1511.05190}}].

\bibitem{jets_w}
P.~Baldi, K.~Bauer, C.~Eng, P.~Sadowski and D.~Whiteson, \emph{{Jet
  substructure classification in high-energy physics with deep neural
  networks}}, \href{https://doi.org/10.1103/PhysRevD.93.094034}{\emph{Phys.
  Rev. D} {\bfseries 93} (2016) 094034}
  [\href{https://arxiv.org/abs/1603.09349}{{\ttfamily 1603.09349}}].

\bibitem{deep_top1}
G.~Kasieczka, T.~Plehn, M.~Russell and T.~Schell, \emph{{Deep-learning Top
  Taggers or The End of QCD?}},
  \href{https://doi.org/10.1007/JHEP05(2017)006}{\emph{JHEP} {\bfseries 1705}
  (2017) 006} [\href{https://arxiv.org/abs/1701.08784}{{\ttfamily
  1701.08784}}].

\bibitem{jets_comparison}
G.~Kasieczka, T.~Plehn et~al., \emph{{The Machine Learning Landscape of Top
  Taggers}}, \href{https://doi.org/10.21468/SciPostPhys.7.1.014}{\emph{SciPost
  Phys.} {\bfseries 7} (2019) 014}
  [\href{https://arxiv.org/abs/1902.09914}{{\ttfamily 1902.09914}}].

\bibitem{particlenet}
H.~Qu and L.~Gouskos, \emph{{ParticleNet: Jet Tagging via Particle Clouds}},
  \href{https://doi.org/10.1103/physrevd.101.056019}{\emph{Physical Review D}
  {\bfseries 101} (2020) } [\href{https://arxiv.org/abs/1902.08570}{{\ttfamily
  1902.08570}}].

\bibitem{jedinet}
E.~A. Moreno, O.~Cerri, J.~M. Duarte, H.~B. Newman, T.~Q. Nguyen, A.~Periwal
  et~al., \emph{{JEDI-net: a jet identification algorithm based on interaction
  networks}}, \href{https://doi.org/10.1140/epjc/s10052-020-7608-4}{\emph{Eur.
  Phys. J.} {\bfseries C80} (2020) 58}
  [\href{https://arxiv.org/abs/1908.05318}{{\ttfamily 1908.05318}}].

\bibitem{DeepJet}
{CMS Collaboration}, \emph{{Performance of the DeepJet b tagging algorithm
  using 41.9/fb of data from proton-proton collisions at 13TeV with Phase 1 CMS
  detector}}, {\emph{CMS-DP-2018-058} (2018) }.

\bibitem{Du:2021pqa}
Y.-L. Du, D.~Pablos and K.~Tywoniuk, \emph{{Jet tomography in heavy ion
  collisions with deep learning}},
  \href{https://arxiv.org/abs/2106.11271}{{\ttfamily 2106.11271}}.

\bibitem{Du:2020pmp}
Y.-L. Du, D.~Pablos and K.~Tywoniuk, \emph{{Deep learning jet modifications in
  heavy-ion collisions}},
  \href{https://doi.org/10.1007/JHEP03(2021)206}{\emph{JHEP} {\bfseries 21}
  (2020) 206} [\href{https://arxiv.org/abs/2012.07797}{{\ttfamily
  2012.07797}}].

\bibitem{Apolinario:2021olp}
L.~Apolin\'ario, N.~F. Castro, M.~Crispim Rom\~ao, J.~G. Milhano, R.~Pedro and
  F.~C.~R. Peres, \emph{{Deep Learning for the classification of quenched
  jets}}, \href{https://doi.org/10.1007/JHEP11(2021)219}{\emph{JHEP} {\bfseries
  11} (2021) 219} [\href{https://arxiv.org/abs/2106.08869}{{\ttfamily
  2106.08869}}].

\bibitem{Berger:2003iw}
C.~F. Berger, T.~Kucs and G.~F. Sterman, \emph{{Event shape / energy flow
  correlations}},
  \href{https://doi.org/10.1103/PhysRevD.68.014012}{\emph{Phys.Rev.} {\bfseries
  D68} (2003) 014012} [\href{https://arxiv.org/abs/hep-ph/0303051}{{\ttfamily
  hep-ph/0303051}}].

\bibitem{Almeida:2008yp}
L.~G. Almeida, S.~J. Lee, G.~Perez, G.~F. Sterman, I.~Sung and J.~Virzi,
  \emph{{Substructure of high-$p_T$ Jets at the LHC}},
  \href{https://doi.org/10.1103/PhysRevD.79.074017}{\emph{Phys. Rev.}
  {\bfseries D79} (2009) 074017}
  [\href{https://arxiv.org/abs/0807.0234}{{\ttfamily 0807.0234}}].

\bibitem{Larkoski:2013eya}
A.~J. Larkoski, G.~P. Salam and J.~Thaler, \emph{{Energy Correlation Functions
  for Jet Substructure}},
  \href{https://doi.org/10.1007/JHEP06(2013)108}{\emph{JHEP} {\bfseries 1306}
  (2013) 108} [\href{https://arxiv.org/abs/1305.0007}{{\ttfamily 1305.0007}}].

\bibitem{Field:1977fa}
R.~D. Field and R.~P. Feynman, \emph{{A parametrization of the properties of
  quark jets}}, \href{https://doi.org/10.1016/0550-3213(78)90015-9}{\emph{Nucl.
  Phys.} {\bfseries B136} (1978) 1}.

\bibitem{Krohn:2012fg}
D.~Krohn, M.~D. Schwartz, T.~Lin and W.~J. Waalewijn, \emph{{Jet Charge at the
  LHC}}, \href{https://doi.org/10.1103/PhysRevLett.110.212001}{\emph{Phys. Rev.
  Lett.} {\bfseries 110} (2013) 212001}
  [\href{https://arxiv.org/abs/1209.2421}{{\ttfamily 1209.2421}}].

\bibitem{Kang:2021ryr}
Z.-B. Kang, X.~Liu, S.~Mantry, M.~C. Spraker and T.~Wilson, \emph{{Dynamic Jet
  Charge}}, \href{https://doi.org/10.1103/PhysRevD.103.074028}{\emph{Phys. Rev.
  D} {\bfseries 103} (2021) 074028}
  [\href{https://arxiv.org/abs/2101.04304}{{\ttfamily 2101.04304}}].

\bibitem{Frye:2017yrw}
C.~Frye, A.~J. Larkoski, J.~Thaler and K.~Zhou, \emph{{Casimir Meets Poisson:
  Improved Quark/Gluon Discrimination with Counting Observables}},
  \href{https://doi.org/10.1007/JHEP09(2017)083}{\emph{JHEP} {\bfseries 09}
  (2017) 083} [\href{https://arxiv.org/abs/1704.06266}{{\ttfamily
  1704.06266}}].

\bibitem{Louppe:2017ipp}
G.~Louppe, K.~Cho, C.~Becot and K.~Cranmer, \emph{{QCD-Aware Recursive Neural
  Networks for Jet Physics}},
  \href{https://doi.org/10.1007/JHEP01(2019)057}{\emph{JHEP} {\bfseries 01}
  (2019) 057} [\href{https://arxiv.org/abs/1702.00748}{{\ttfamily
  1702.00748}}].

\bibitem{Komiske:2018cqr}
P.~T. Komiske, E.~M. Metodiev and J.~Thaler, \emph{{Energy Flow Networks: Deep
  Sets for Particle Jets}},
  \href{https://doi.org/10.1007/JHEP01(2019)121}{\emph{JHEP} {\bfseries 01}
  (2019) 121} [\href{https://arxiv.org/abs/1810.05165}{{\ttfamily
  1810.05165}}].

\bibitem{Qu:2019gqs}
H.~Qu and L.~Gouskos, \emph{{ParticleNet: Jet Tagging via Particle Clouds}},
  \href{https://doi.org/10.1103/PhysRevD.101.056019}{\emph{Phys. Rev. D}
  {\bfseries 101} (2020) 056019}
  [\href{https://arxiv.org/abs/1902.08570}{{\ttfamily 1902.08570}}].

\bibitem{Lee:2019ssx}
J.~S.~H. Lee, S.~M. Lee, Y.~Lee, I.~Park, I.~J. Watson and S.~Yang,
  \emph{{Quark Gluon Jet Discrimination with Weakly Supervised Learning}},
  \href{https://doi.org/10.3938/jkps.75.652}{\emph{J. Korean Phys. Soc.}
  {\bfseries 75} (2019) 652}
  [\href{https://arxiv.org/abs/2012.02540}{{\ttfamily 2012.02540}}].

\bibitem{Dreyer:2020brq}
F.~A. Dreyer and H.~Qu, \emph{{Jet tagging in the Lund plane with graph
  networks}}, \href{https://doi.org/10.1007/JHEP03(2021)052}{\emph{JHEP}
  {\bfseries 03} (2021) 052}
  [\href{https://arxiv.org/abs/2012.08526}{{\ttfamily 2012.08526}}].

\bibitem{Metodiev:2018ftz}
E.~M. Metodiev and J.~Thaler, \emph{{Jet Topics: Disentangling Quarks and
  Gluons at Colliders}},
  \href{https://doi.org/10.1103/PhysRevLett.120.241602}{\emph{Phys. Rev. Lett.}
  {\bfseries 120} (2018) 241602}
  [\href{https://arxiv.org/abs/1802.00008}{{\ttfamily 1802.00008}}].

\bibitem{Brewer:2020och}
J.~Brewer, J.~Thaler and A.~P. Turner, \emph{{Data-driven quark and gluon jet
  modification in heavy-ion collisions}},
  \href{https://doi.org/10.1103/PhysRevC.103.L021901}{\emph{Phys. Rev. C}
  {\bfseries 103} (2021) L021901}
  [\href{https://arxiv.org/abs/2008.08596}{{\ttfamily 2008.08596}}].

\bibitem{Dreyer:2018nbf}
F.~A. Dreyer, G.~P. Salam and G.~Soyez, \emph{{The Lund Jet Plane}},
  \href{https://doi.org/10.1007/JHEP12(2018)064}{\emph{JHEP} {\bfseries 12}
  (2018) 064} [\href{https://arxiv.org/abs/1807.04758}{{\ttfamily
  1807.04758}}].

\bibitem{Aad:2020zcn}
{\scshape ATLAS} collaboration, \emph{{Measurement of the Lund Jet Plane Using
  Charged Particles in 13 TeV Proton-Proton Collisions with the ATLAS
  Detector}}, \href{https://doi.org/10.1103/PhysRevLett.124.222002}{\emph{Phys.
  Rev. Lett.} {\bfseries 124} (2020) 222002}
  [\href{https://arxiv.org/abs/2004.03540}{{\ttfamily 2004.03540}}].

\bibitem{ALICE-PUBLIC-2021-002}
{\scshape ALICE Collaboration} collaboration, \emph{{Physics Preliminary
  Summary: Measurement of the primary Lund plane density in pp collisions at
  $\sqrt{s} = \rm{13}$ TeV with ALICE}}, .

\bibitem{Lifson:2020gua}
A.~Lifson, G.~P. Salam and G.~Soyez, \emph{{Calculating the primary Lund Jet
  Plane density}}, \href{https://doi.org/10.1007/JHEP10(2020)170}{\emph{JHEP}
  {\bfseries 10} (2020) 170}
  [\href{https://arxiv.org/abs/2007.06578}{{\ttfamily 2007.06578}}].

\bibitem{Soper:2011cr}
D.~E. Soper and M.~Spannowsky, \emph{{Finding physics signals with shower
  deconstruction}},
  \href{https://doi.org/10.1103/PhysRevD.84.074002}{\emph{Phys. Rev.}
  {\bfseries D84} (2011) 074002}
  [\href{https://arxiv.org/abs/1102.3480}{{\ttfamily 1102.3480}}].

\bibitem{Soper:2012pb}
D.~E. Soper and M.~Spannowsky, \emph{{Finding top quarks with shower
  deconstruction}},
  \href{https://doi.org/10.1103/PhysRevD.87.054012}{\emph{Phys. Rev.}
  {\bfseries D87} (2013) 054012}
  [\href{https://arxiv.org/abs/1211.3140}{{\ttfamily 1211.3140}}].

\bibitem{FerreiradeLima:2016gcz}
D.~Ferreira~de Lima, P.~Petrov, D.~Soper and M.~Spannowsky, \emph{{Quark-Gluon
  tagging with Shower Deconstruction: Unearthing dark matter and Higgs
  couplings}}, \href{https://doi.org/10.1103/PhysRevD.95.034001}{\emph{Phys.
  Rev.} {\bfseries D95} (2017) 034001}
  [\href{https://arxiv.org/abs/1607.06031}{{\ttfamily 1607.06031}}].

\bibitem{Dokshitzer:1997in}
Y.~L. Dokshitzer, G.~Leder, S.~Moretti and B.~Webber, \emph{{Better jet
  clustering algorithms}},
  \href{https://doi.org/10.1088/1126-6708/1997/08/001}{\emph{JHEP} {\bfseries
  9708} (1997) 001} [\href{https://arxiv.org/abs/hep-ph/9707323}{{\ttfamily
  hep-ph/9707323}}].

\bibitem{Wobisch:1998wt}
M.~Wobisch and T.~Wengler, \emph{{Hadronization corrections to jet
  cross-sections in deep inelastic scattering}},
  \href{https://arxiv.org/abs/hep-ph/9907280}{{\ttfamily hep-ph/9907280}}.

\bibitem{Cacciari:2008gp}
M.~Cacciari, G.~P. Salam and G.~Soyez, \emph{{The Anti-k(t) jet clustering
  algorithm}}, \href{https://doi.org/10.1088/1126-6708/2008/04/063}{\emph{JHEP}
  {\bfseries 04} (2008) 063} [\href{https://arxiv.org/abs/0802.1189}{{\ttfamily
  0802.1189}}].

\bibitem{Ehlers:2021njb}
{\scshape ALICE} collaboration, \emph{{QCD dynamics studied with jets in
  ALICE}},  in \emph{{55th Rencontres de Moriond on QCD and High Energy
  Interactions}}, 5, 2021, \href{https://arxiv.org/abs/2105.10523}{{\ttfamily
  2105.10523}}.

\bibitem{Larkoski:2014wba}
A.~J. Larkoski, S.~Marzani, G.~Soyez and J.~Thaler, \emph{{Soft Drop}},
  \href{https://doi.org/10.1007/JHEP05(2014)146}{\emph{JHEP} {\bfseries 1405}
  (2014) 146} [\href{https://arxiv.org/abs/1402.2657}{{\ttfamily 1402.2657}}].

\bibitem{Dasgupta:2013ihk}
M.~Dasgupta, A.~Fregoso, S.~Marzani and G.~P. Salam, \emph{{Towards an
  understanding of jet substructure}},
  \href{https://doi.org/10.1007/JHEP09(2013)029}{\emph{JHEP} {\bfseries 1309}
  (2013) 029} [\href{https://arxiv.org/abs/1307.0007}{{\ttfamily 1307.0007}}].

\bibitem{Dasgupta:2001sh}
M.~Dasgupta and G.~Salam, \emph{{Resummation of nonglobal QCD observables}},
  \href{https://doi.org/10.1016/S0370-2693(01)00725-0}{\emph{Phys.Lett.}
  {\bfseries B512} (2001) 323}
  [\href{https://arxiv.org/abs/hep-ph/0104277}{{\ttfamily hep-ph/0104277}}].

\bibitem{Hatta:2013iba}
Y.~Hatta and T.~Ueda, \emph{{Resummation of non-global logarithms at finite
  $N_c$}},
  \href{https://doi.org/10.1016/j.nuclphysb.2013.06.021}{\emph{Nucl.Phys.}
  {\bfseries B874} (2013) 808}
  [\href{https://arxiv.org/abs/1304.6930}{{\ttfamily 1304.6930}}].

\bibitem{Hamilton:2020rcu}
K.~Hamilton, R.~Medves, G.~P. Salam, L.~Scyboz and G.~Soyez, \emph{{Colour and
  logarithmic accuracy in final-state parton showers}},
  \href{https://doi.org/10.1007/JHEP03(2021)041}{\emph{JHEP} {\bfseries 03}
  (2021) 041} [\href{https://arxiv.org/abs/2011.10054}{{\ttfamily
  2011.10054}}].

\bibitem{lstm1997}
S.~Hochreiter and J.~Schmidhuber, \emph{Long short-term memory}, {\emph{Neural
  computation} {\bfseries 9} (1997) 1735}.

\bibitem{DBLP:journals/corr/HeZR015}
K.~He, X.~Zhang, S.~Ren and J.~Sun, \emph{Delving deep into rectifiers:
  Surpassing human-level performance on imagenet classification}, {\emph{CoRR}
  {\bfseries abs/1502.01852} (2015) }
  [\href{https://arxiv.org/abs/1502.01852}{{\ttfamily 1502.01852}}].

\bibitem{DBLP:journals/corr/KingmaB14}
D.~P. Kingma and J.~Ba, \emph{Adam: {A} method for stochastic optimization},
  {\emph{CoRR} {\bfseries abs/1412.6980} (2014) }
  [\href{https://arxiv.org/abs/1412.6980}{{\ttfamily 1412.6980}}].

\bibitem{lundnet_code}
F.~A. Dreyer and H.~Qu, ``{LundNet} v1.0.0.''
  \url{https://doi.org/10.5281/zenodo.4443152}.
\newblock 10.5281/zenodo.4443152.

\bibitem{DGCNN}
Y.~Wang, Y.~Sun, Z.~Liu, S.~E. Sarma, M.~M. Bronstein and J.~M. Solomon,
  \emph{Dynamic graph cnn for learning on point clouds},
  \href{https://doi.org/10.1145/3326362}{\emph{ACM Trans. Graph.} {\bfseries
  38} (2019) 146}.

\bibitem{DBLP:journals/corr/IoffeS15}
S.~Ioffe and C.~Szegedy, \emph{Batch normalization: Accelerating deep network
  training by reducing internal covariate shift},  in \emph{Proceedings of the
  32nd International Conference on Machine Learning}, vol.~37, (Lille, France),
  pp.~448--456, PMLR, 07--09 Jul, 2015,
  \href{http://proceedings.mlr.press/v37/ioffe15.html}{http://proceedings.mlr.press/v37/ioffe15.html}.

\bibitem{glorot2011deep}
X.~Glorot, A.~Bordes and Y.~Bengio, \emph{Deep sparse rectifier neural
  networks},  in \emph{Proceedings of the Fourteenth International Conference
  on Artificial Intelligence and Statistics}, vol.~15, (Fort Lauderdale, FL,
  USA), pp.~315--323, PMLR, 11--13 Apr, 2011,
  \href{http://proceedings.mlr.press/v15/glorot11a.html}{http://proceedings.mlr.press/v15/glorot11a.html}.

\bibitem{he2016deep}
K.~{He}, X.~{Zhang}, S.~{Ren} and J.~{Sun}, \emph{Deep residual learning for
  image recognition},  in \emph{2016 IEEE Conference on Computer Vision and
  Pattern Recognition (CVPR)}, (Las Vegas, NV, USA), pp.~770--778, IEEE, 2016,
  \href{https://doi.org/10.1109/CVPR.2016.90}{DOI}.

\bibitem{wang2020deep}
M.~Wang, D.~Zheng, Z.~Ye, Q.~Gan, M.~Li, X.~Song et~al., \emph{Deep graph
  library: A graph-centric, highly-performant package for graph neural
  networks},  2020.

\bibitem{NEURIPS2019_9015}
A.~Paszke, S.~Gross, F.~Massa, A.~Lerer, J.~Bradbury, G.~Chanan et~al.,
  \emph{Pytorch: An imperative style, high-performance deep learning library},
  in \emph{Advances in Neural Information Processing Systems 32}, H.~Wallach,
  H.~Larochelle, A.~Beygelzimer, F.~d\textquotesingle Alch\'{e}-Buc, E.~Fox and
  R.~Garnett, eds., pp.~8024--8035, Curran Associates, Inc., (2019),
  \href{http://papers.neurips.cc/paper/9015-pytorch-an-imperative-style-high-performance-deep-learning-library.pdf}{http://papers.neurips.cc/paper/9015-pytorch-an-imperative-style-high-performance-deep-learning-library.pdf}.

\bibitem{Dasgupta:2014yra}
M.~Dasgupta, F.~Dreyer, G.~P. Salam and G.~Soyez, \emph{{Small-radius jets to
  all orders in QCD}},
  \href{https://doi.org/10.1007/JHEP04(2015)039}{\emph{JHEP} {\bfseries 04}
  (2015) 039} [\href{https://arxiv.org/abs/1411.5182}{{\ttfamily 1411.5182}}].

\bibitem{Sjostrand:2006za}
T.~Sj{\"o}strand, S.~Mrenna and P.~Skands, \emph{{PYTHIA 6.4 physics and
  manual}}, {\emph{JHEP} {\bfseries 05} (2006) 026}
  [\href{https://arxiv.org/abs/hep-ph/0603175}{{\ttfamily hep-ph/0603175}}].

\bibitem{Sjostrand:2014zea}
T.~Sjöstrand, S.~Ask, J.~R. Christiansen, R.~Corke, N.~Desai, P.~Ilten et~al.,
  \emph{{An Introduction to PYTHIA 8.2}},
  \href{https://doi.org/10.1016/j.cpc.2015.01.024}{\emph{Comput. Phys. Commun.}
  {\bfseries 191} (2015) 159}
  [\href{https://arxiv.org/abs/1410.3012}{{\ttfamily 1410.3012}}].

\bibitem{Skands:2014pea}
P.~Skands, S.~Carrazza and J.~Rojo, \emph{{Tuning PYTHIA 8.1: the Monash 2013
  Tune}}, \href{https://doi.org/10.1140/epjc/s10052-014-3024-y}{\emph{Eur.
  Phys. J.} {\bfseries C74} (2014) 3024}
  [\href{https://arxiv.org/abs/1404.5630}{{\ttfamily 1404.5630}}].

\bibitem{Cacciari:2005hq}
M.~Cacciari and G.~P. Salam, \emph{{Dispelling the $N^{3}$ myth for the $k_t$
  jet-finder}},
  \href{https://doi.org/10.1016/j.physletb.2006.08.037}{\emph{Phys. Lett.}
  {\bfseries B641} (2006) 57}
  [\href{https://arxiv.org/abs/hep-ph/0512210}{{\ttfamily hep-ph/0512210}}].

\bibitem{Cacciari:2011ma}
M.~Cacciari, G.~P. Salam and G.~Soyez, \emph{{FastJet User Manual}},
  \href{https://doi.org/10.1140/epjc/s10052-012-1896-2}{\emph{Eur. Phys. J.}
  {\bfseries C72} (2012) 1896}
  [\href{https://arxiv.org/abs/1111.6097}{{\ttfamily 1111.6097}}].

\bibitem{Bahr:2008pv}
M.~Bahr et~al., \emph{{Herwig++ Physics and Manual}},
  \href{https://doi.org/10.1140/epjc/s10052-008-0798-9}{\emph{Eur. Phys. J.}
  {\bfseries C58} (2008) 639}
  [\href{https://arxiv.org/abs/0803.0883}{{\ttfamily 0803.0883}}].

\bibitem{Bellm:2015jjp}
J.~Bellm et~al., \emph{{Herwig 7.0/Herwig++ 3.0 release note}},
  \href{https://doi.org/10.1140/epjc/s10052-016-4018-8}{\emph{Eur. Phys. J. C}
  {\bfseries 76} (2016) 196}
  [\href{https://arxiv.org/abs/1512.01178}{{\ttfamily 1512.01178}}].

\bibitem{particle-net-code}
H.~Qu and L.~Gouskos, ``Particle-net.'' https://github.com/hqucms/ParticleNet,
  (as of 2021-06-21).

\bibitem{energy-flow-code}
P.~Komiske, E.~Metodiev and J.~Thaler, ``Energy flow.''
  https://energyflow.network, (as of 2021-06-21).

\bibitem{Dasgupta:2020fwr}
M.~Dasgupta, F.~A. Dreyer, K.~Hamilton, P.~F. Monni, G.~P. Salam and G.~Soyez,
  \emph{{Parton showers beyond leading logarithmic accuracy}},
  \href{https://doi.org/10.1103/PhysRevLett.125.052002}{\emph{Phys. Rev. Lett.}
  {\bfseries 125} (2020) 052002}
  [\href{https://arxiv.org/abs/2002.11114}{{\ttfamily 2002.11114}}].

\bibitem{Bright-Thonney:2018mxq}
S.~Bright-Thonney and B.~Nachman, \emph{{Investigating the Topology Dependence
  of Quark and Gluon Jets}},
  \href{https://doi.org/10.1007/JHEP03(2019)098}{\emph{JHEP} {\bfseries 03}
  (2019) 098} [\href{https://arxiv.org/abs/1810.05653}{{\ttfamily
  1810.05653}}].

\end{thebibliography}\endgroup

\end{document}